\documentclass[doublespacing]{elsart}
\usepackage{graphicx}
\usepackage{dcolumn}
\usepackage{bm}
\usepackage{amssymb}
\makeatletter
\DeclareRobustCommand*\cal{\@fontswitch\relax\mathcal}
\makeatother
\begin{document}

\begin{frontmatter}
\title{A General Theory of Goodness of Fit in Likelihood Fits}
\author{Rajendran Raja}
 \ead{raja@fnal.gov}
\address{%
Fermi National Accelerator Laboratory\\
Batavia, IL 60510}

\begin{abstract}
Maximum likelihood fits to data can be performed using binned data and
unbinned data. The likelihood fits in either case produce only the
fitted quantities but not the goodness of fit.  With binned data, one
can obtain a measure of the goodness of fit by using the $\chi^2$
method, after the maximum likelihood fitting is performed. With
unbinned data, currently, the fitted parameters are obtained but no
measure of goodness of fit is available. This remains, to date, an
unsolved problem in statistics. By considering the transformation
properties of likelihood functions with respect to change of variable,
we conclude that the likelihood ratio of the theoretically predicted
probability density to that of {\it the data density} is invariant
under change of variable and provides the goodness of fit. We show how
to apply this likelihood ratio for binned as well as unbinned
likelihoods and show that even the $\chi^2$ test is a special case of
this general theory. In order to calculate errors in the fitted
quantities, we need to solve the problem of inverse probabilities. We
use Bayes' theorem to do this, using the data density obtained  in the
goodness of fit. This permits one to invert the probabilities without
the use of a Bayesian prior. The resulting statistics is consistent
with frequentist ideas.

\end{abstract}
\end{frontmatter}
\tableofcontents
\newpage
\section{Introduction}
In particle physics as well as other branches of science, fitting
theoretical models to data is a crucial end stage to the performance
of experiments. Minimizing the $\chi^2$ between theory and experiment
is perhaps the most commonly used form of fitting, with data binned in
histograms. Such fits yield not only the fitted parameters and errors
on the fitted parameters but also a measure of the goodness of fit.
Another common fitting method is the maximum likelihood method which
can be performed on binned and unbinned data to obtain the best values
of theoretical parameters. In the case of unbinned likelihood fitting, there
is currently no measure of the goodness of fit. In this paper, we
propose a solution to the problem, which by its nature works generally
for both binned and unbinned likelihood fits. A general theory of goodness of
fit in likelihood fits results.

\subsection{Notation}
In what follows, we will denote by the vector $s$, the theoretical
parameters ($s$ for ``signal'') and the vector $c$, the experimentally
measured quantities or ``configurations''. For simplicity, we will
illustrate the method where both $s$ and $c$ are one dimensional,
though either or both can be multi-dimensional in practice. We thus
define the theoretical model by the conditional probability density
$P(c|s)$, defined as the probability of observing $c$ given a value of $s$.
The theoretical probability function obeys the normalization condition
\begin{equation}
\int  P(c|s) dc = 1
\label{thenorm}
\end{equation}
Then an unbinned maximum likelihood fit to data is obtained
by maximizing the likelihood~\cite{fisher},
\begin{equation}
{\cal L} = \prod_{i=1}^{i=n} P(c_i|s)
\label{like}
\end{equation}
where the likelihood is evaluated at the $n$ observed data points
$c_i, i=1,n$.  Such a fit will determine the maximum likelihood value
$s^*$ of the theoretical parameters, but will not tell us how good the
fit is.
\subsection{To show that ${\cal L}$ cannot be used as a goodness of fit 
variable}

The goodness of fit variable must be invariant under a change of
variable $c \rightarrow c'$.  The value of the likelihood ${\cal L}$
at the maximum likelihood point does not furnish a goodness of fit,
since the likelihood is not invariant under change of variable. This
can be seen by observing that one can transform the variable set $c$
to a variable set $c'$ such that $P(c'|s^*)$ is uniformly distributed
between 0 and 1. In one dimension, this is trivially done by the
transformation function $c'(c)$ such that

\begin{equation}
  c'(c) = \int_{c_1}^{c} P(t|s^*) dt 
\label{hyp}
\end{equation}
The variable $c$ ranges from $c_1$ to $c_2$ and the 
probability function $P(c|s^*)$ normalizes to unity in this range. 
This implies that $c'$ ranges from 0 to 1.
Such a transformation is known as a hypercube transformation, in
multi-dimensions. The transformed probability distribution in the variable $c'$ is unity in this interval as can be seen by examining the Jacobian of the 
transformation $|\frac{\partial c'}{\partial c}|$
\begin{eqnarray}
 |\frac{\partial c'}{\partial c}| = P(c|s^*)\\
 P(c'|s^*) = P(c|s^*)|\frac{\partial c}{\partial c'}| = 1
\end{eqnarray}
Other datasets will yield different values of
likelihood in the variable space $c$ when the likelihood is computed
with the original function $P(c|s^*)$. However, in 
hypercube space, the value of the likelihood is unity regardless of
the dataset $c'_i,i=1,n$, thus the likelihood ${\cal L}$ cannot
furnish a goodness of fit by itself, since neither the likelihood, nor
ratios of likelihoods computed using the same distribution $P(c|s^*)$
is invariant under variable transformations. The fundamental reason
for this non-invariance is that only a single distribution, namely,
$P(c|s^*)$ is being used to compute the goodness of fit.

\begin{figure}[tbh!]
\begin{minipage}{15pc}
\includegraphics[width=15pc]{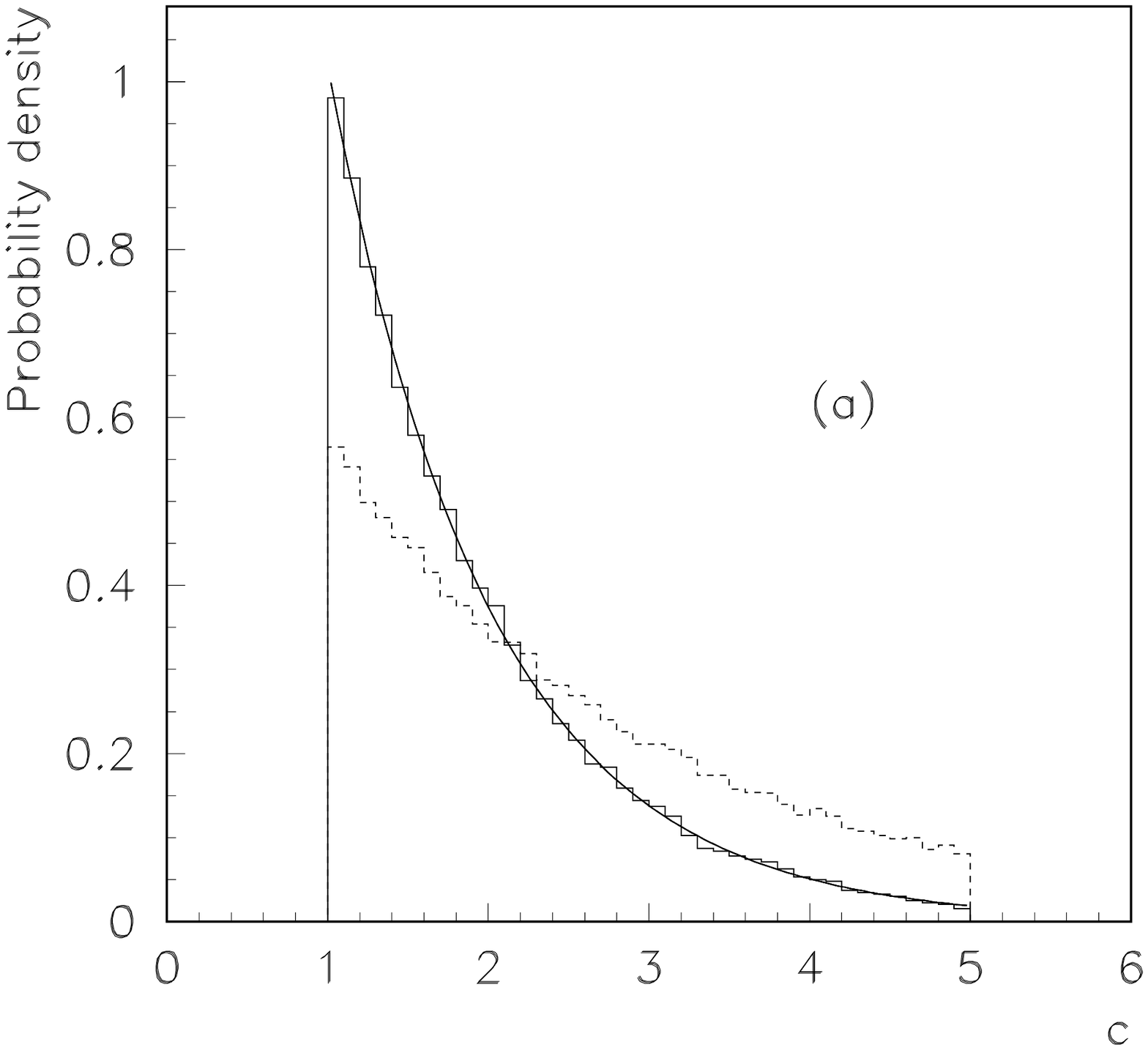}
\end{minipage}
\begin{minipage}{15pc}
\includegraphics[width=15pc]{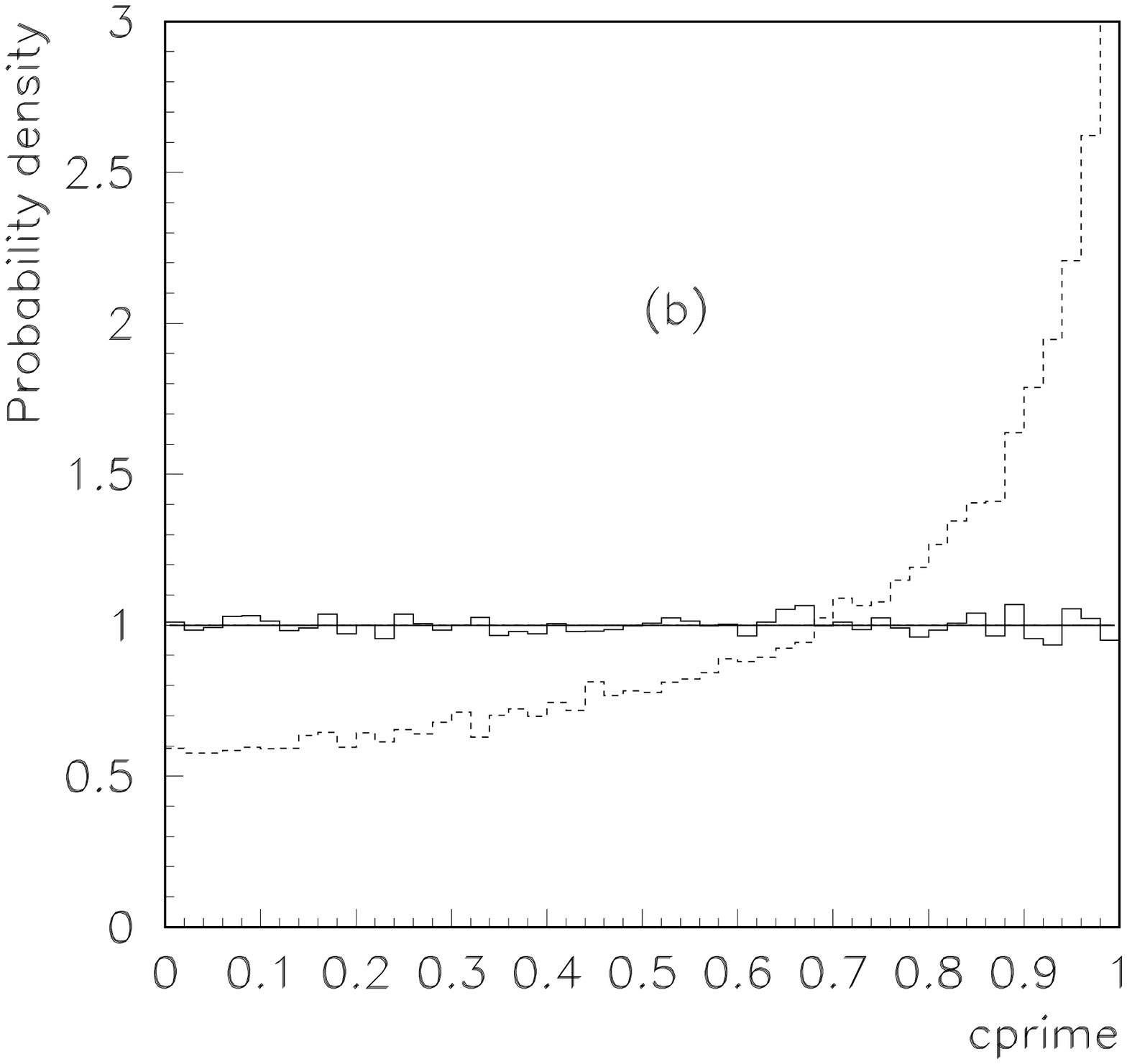}
\end{minipage}
\caption
{(a) shows the fitting in the dataset space. The curve shows the
fitted function. Superimposed is the fitted data, (full histogram,
normalized to unity). The dashed histogram shows the different dataset
which obviously does not fit to the fitted curve. (b) The same plot in
hyperspace. The fitted function is flat by construction. Both the
fitted data set (full histogram) and the dashed histogram will have the
same value of likelihood ${\cal L}$ in this space which implies that
${\cal L}$ cannot be used as a goodness of fit variable.
\label{exps}}
\end{figure}

To illustrate further, we use a concrete example of fitting a dataset
using the maximum likelihood method as shown in Figure~\ref{exps}(a). 
The fitting is done in the range $c_1<c<c_2$, 
where $c_1=1.0$ and $c_2=5.0$. The fitting function is
\begin{equation}
 P(c|s) = \frac{\exp (-c/s)}{ s(\exp (-c_1/s) - \exp(-c_2/s))}
\end{equation}
which normalizes to unity in the range $c_1<c<c_2$.
The fitted dataset is shown as a full histogram. The dashed histogram
shows a dataset that is a poor fit to the data and will produce a
smaller value of ${\cal L}$ when fitted as a function of
$c$. Figure~\ref{exps}(b) shows the same data in the hypercube space
where the fitted function is flat as per the transformation given in
equation~\ref{hyp}. Both the datasets will produce a value of unity for
${\cal L}$ in this space implying an equally good fit in either case,
which is obviously false. This clearly demonstrates that the likelihood 
by itself 
cannot provide a goodness of fit variable.
\section{Likelihood ratios}
\label{lrat}

\subsection{The concept of ``data likelihood'' derived from 
the {\it pdf} of the data}

It is interesting to note that while using $\chi^2$ as the goodness of
fit technique for binned histograms, we use two distribution
functions, namely the theoretical curve and the data. By binning the
data, we are in effect estimating the probability density function of
the data as the second distribution, in addition to the theoretical 
distribution specified by the theoretical curve. 
In likelihood language we define the probability density function ($pdf$) 
of  the data as 
\begin{equation}
  P^{data}(c) = \lim_{N\rightarrow \infty} \frac{1}{N} \frac{dN}{dc}
\end{equation}
where $N$ is the number of times the experiment is repeated that 
results in the observable $c$. The function $P^{data}(c)$ obeys 
the normalization condition
\begin{equation}
 \int P^{data}(c) dc = 1
\end{equation}
When one is using binned likelihoods, the $pdf$ of the data would be
estimated by binning the events in a histogram and normalizing the sum
of contents of all bins to unity.  In the unbinned case, we
will describe below a technique~\cite{parzen} on estimating $P^{data}(c)$ using
Probability Density Estimators ($PDE$).

We can now define a likelihood ratio $\cal L_R$ such that
\begin{equation}
{\cal L_R} = \frac{\prod_{i=1}^{i=n} P(c_i|s)}{\prod_{i=1}^{i=n} P^{data}(c_i)}
\equiv \frac{P(\vec{c_n}|s)}{P^{data}(\vec{c_n})}
\label{lrd}
\end{equation}
where we have used the notation $\vec{c_n}$ to denote 
the  dataset $c_i, i=1,n$.

Since the $n$ events $c_i,i=1,n$ are independent, the probability of obtaining
the dataset $\vec{c_n}$ is given by
\begin{equation}
  P^{data}(\vec{c_n}) = \prod_{i=1}^{i=n}P^{data}(c_i)
\end{equation}
The quantity $ P^{data}(\vec{c_n})$ we 
name the ``data likelihood'' of the dataset $\vec{c_n}$ and the 
quantity $P(\vec{c_n}|s)$ as the ``theory likelihood'' of the 
dataset $\vec{c_n}$. We note that the ``data likelihood'' 
$ P^{data}(\vec{c_n})$ may also be thought of as the probability density of
the`` $n-object$'' $\vec{c_n}$ which obeys the normalization condition
\begin{equation}
 \int P^{data}(\vec{c_n})~d\vec{c_n} = 1
\end{equation}
Let us now note that ${\cal L_R}$ is invariant 
under a general variable transformation (not restricted to hypercube 
transformation)
$c\rightarrow c'$, since 
\begin{eqnarray}
 P(c'|s) = |\frac{\partial c}{\partial c'}|P(c|s) \\
 P^{data}(c') = |\frac{\partial c}{\partial c'}|P^{data}(c) \\
 {\cal L'_R} = {\cal L_R}
\end{eqnarray}
and the Jacobian of the transformation $|\frac{\partial c}{\partial
c'}|$ cancels in the numerator and denominator in the ratio. This is
an extremely important property of the likelihood ratio ${\cal L_R}$
that qualifies it to be a goodness of fit variable. Later, we will
show that the binned likelihood ratio asymptotically approaches a
$\chi^2$ distribution as the number of events $n\rightarrow \infty$,
further motivating this choice. Since the denominator
$P^{data}(\vec{c_n})$ is independent of the theoretical parameters
$s$, both the likelihood ratio and the likelihood maximize at the same
point $s^*$. The likelihood ratios for two different data sets
$\vec{c_m}$ and $\vec{c_n}$ can be combined by multiplication as per

\begin{equation}
 {\cal L_R}^{m+n} = {\cal L_R}^m\times {\cal L_R}^n
\label{mult}
\end{equation}
This rule follows from the definition of ${\cal L_R}$ in equation~\ref{lrd}.
In practice, we will use the negative log-likelihood ratio
 ${\cal NLLR}=-log_e {\cal L}_R$
as the goodness of fit variable and minimize it. The multiplication rule of 
equation~\ref{mult} results in an addition rule for ${\cal NLLR}$.
The problem of finding the distribution of ${\cal NLLR}$ for a good
fit then reduces to finding the distribution of ${\cal NLLR}$ in
hyper-cube space for a variable that is uniformly distributed between
zero and one, as in Figure~\ref{exps}(b). This is because ${\cal
NLLR}$ is invariant under the transformation of variable. So all
goodness of fit problems using likelihood ratios 
can be reduced to finding the distribution of
${\cal NLLR}$ for a variable that is uniformly distributed in
hypercube space.

\subsection{Historical use of Likelihood Ratios}

The Neyman-Pearson lemma~\cite{neym} states that if one 
is trying to choose between two 
hypotheses $H_0$ and $H_1$, then the cut on the likelihood ratio 
\begin{equation}
 {\cal L}_R = \frac{P(\vec{c_n}|H_0)}{P(\vec{c_n}|H_1)} > \epsilon
\end{equation}
will have the optimum power in differentiating between the hypotheses
$H_0$ and $H_1$, where $\epsilon$ is a constant adjusted to obtain the
desired purity in favor of hypothesis $H_0$. Notice that this
likelihood ratio is between the likelihood computed for two different
hypotheses $H_0$ and $H_1$. Our likelihood ratio differs fundamentally
from this in that the denominator we use $P(\vec{c_n})$ is the ``data
likelihood'' that is computed from the distribution of the data and is
not tied to any hypothesis as such.

\section{Normalizing the theoretical curve to the data}

The method of maximum likelihood fits the shape of the theoretical
distribution to the data distribution. The theoretical model obeys the
normalization condition in equation~\ref{thenorm} and the likelihood
is evaluated at the number of observed data events $n$. There is no
explicit mention of the theoretically expected number of events, which
we denote by $n_t$. Later we will show how to incorporate a goodness of fit 
in the absolute normalization by making use of the binomial distribution 
and its limiting cases the Poisson and the normal distributions. We will begin
by obtaining goodness of fit formulae for the case where we bin the data 
and fit the theoretical shape to the experimental distribution.

\section{Binned Goodness of Fit}

When one  bins  data in histograms and fits the theory shape to the 
data, one can fit by using  either maximum
likelihood or by minimizing $\chi^2$. In either case, the goodness of fit 
is usually evaluated using $\chi^2$. We now illustrate how the likelihood ratio
defined in section~\ref{lrat} can be used to obtain a goodness of fit after 
the maximum likelihood fitting is done.
In order to evaluate the likelihood ratio, one needs to evaluate the
theory likelihood and the data likelihood for each value of $c_i$. For
the binned histogram, we make the approximation of assuming that both
these quantities are constant for all values of $c_i$ in a given bin and
evaluating each at the bin center. Let there be $n_b$ bins and  
let the $k^{th}$ bin contain $n_k$ entries. 
\subsection{The multinomial distribution}

The probability of obtaining the histogram is given by the multinomial
distribution
\begin{eqnarray}
 P(histogram) = \frac{n!}{\prod_{k=1}^{k=n_b} n_k!}
                 \prod_{k=1}^{k=n_b} P(c_k|s)^{n_k}\\
\sum_{k=1}^{k=n_b} n_k = n
\end{eqnarray}
\subsection{Degeneracy of the distribution}

The factor $\frac{n!}{\prod_{k=1}^{k=n_b} n_k!}$ denotes the number of
ways $n$ events can be partitioned to form the observed histogram,
which we term the degeneracy $\cal {D}$ of the histogram. Each of the
${\cal D}$ histograms is identical to each other and possesses the same
goodness of fit.  We can then evaluate the goodness of fit for any one
of the ${\cal D}$ degenerate histograms, the likelihood for which is
given by

\begin{equation}
{\cal L} = \prod_{k=1}^{k=n_b} P(c_k|s)^{n_k}
\end{equation}
and the likelihood ratio can be written as
\begin{equation}
{\cal L}_R = \prod_{k=1}^{k=n_b} \left(\frac{P(c_k|s)}{P^{data}(c_k)}\right) ^{n_k}
\label{lrb}
\end{equation}
The value of $\frac{P(c_k|s)}{P^{data}(c_k))}$  raised to the 
power $n_k$ in equation~\ref{lrb} results from the fact 
that there are $n_k$ configurations 
$c_i$ in the $k^{th}$ bin and we are multiplying 
a constant  ratio (at the bin center) 
over $n_k$ configurations.
If $\Delta c_k$ is the bin width for the $k^{th}$ bin, then 
the data likelihood can be approximated by
\begin{equation}
 P^{data}(c_k) \approx \frac{n_k}{n \Delta c_k} 
\end{equation}
This obeys the normalization condition
\begin{equation}
 \int P^{data}(c_k) dc_k \approx \sum_{k=1}^{k=n_b} \frac{n_k}{n\Delta c_k} \Delta c_k = 1.
\end{equation}
The theoretical likelihood can be integrated over the bin to yield
\begin{equation}
 P^{bin\:average}(c_k|s) = \frac{1}{\Delta c_k}
\int_{c=c_k-\Delta c_k/2}^{c=c_k + \Delta c_k/2}
P(c|s) dc
\label{tint}
\end{equation}
This obeys the normalization condition
\begin{equation}
  \sum_{k=1}^{k=n_b} P^{bin\:average}(c_k|s) \Delta c_k = 1
\label{tnorm}
\end{equation}
Then the likelihood ratio can be written
\begin{equation}
{\cal L_R} = \prod_{k=1}^{k=n_b} \left( \frac {n \Delta c_k P^{bin\:average}(c_k|s)}{n_k}\right)^{n_k} 
\equiv 
\prod_{k=1}^{k=n_b} \left ( \frac {T_k}{n_k}\right)^{n_k} 
\label{lrat2}
\end{equation}
where $T_k \equiv n \Delta c_k P^{bin\:average}(c_k|s)$ is the
theoretically expected number of events in the $k^{th}$ bin obeying
the normalization condition $\sum_k T_k = n$, as per
equation~\ref{tnorm}.  This likelihood ratio may be used to
obtain a maximum likelihood fit as well as to obtain a goodness of
fit. Note that the likelihood ratio is well-behaved even for empty bins 
where $n_k=0$, since $n_k^{n_k}$ is unity for such cases.

Note that the negative log-likelihood ratio ${\cal NLLR}$ resulting from
equation~\ref{lrat2}
yields
\begin{equation}
 {\cal NLLR} = \sum_{k=1}^{k=n_b} n_k \:log_e\:(\frac{n_k}{T_k})
\label{lrat3}
\end{equation}
which is the same result as derived by Baker and Cousins
~\cite{baker} for the multinomial case where normalization is
preserved between theory and experiment. We have derived the result
using very different arguments (than Baker and Cousins) 
for the denominator of the likelihood
ratio, namely it is the value of the $data~pdf$ at the bin center 
as a result of the general theory developed here. 

If we are reluctant to work out (for reasons of computing speed) the
integral in equation~\ref{tint} for each bin at each step of the 
fitting process, then we can approximate it by the bin center values
\begin{equation}
 P^{bin\:average}(c_k|s) \approx \frac{P(c_k|s)}{\sum_k P(c_k|s)\:\Delta c_k }
\end{equation}
This then obeys the normalization equation~\ref{tnorm} and 
the expression in equation~\ref{lrat3} for ${\cal NLLR}$ can be used generally.

\subsection{To Show that the Binned Negative Log-Likelihood Ratio Approaches a 
$\chi^2$ Distribution for Large $n$}
Let the difference between $n_k$, the observed  number of events and 
$T_k$ the  theoretical number of events  
be denoted by $\lambda_k = n_k - T_k$. Then $\sum_k \lambda_k = 0$, 
by virtue of the normalization conditions. Then the binned negative log 
likelihood ratio ${\cal NLLR}$ can be written
\begin{equation}
 {\cal NLLR} = -log_e~{\cal L}_R =  -\sum_{k=1}^{k=n_b} n_k\: log_e \left(1-\frac{\lambda_k}{n_k}\right)
\end{equation}
This can be expanded in powers of $\lambda_k/n_k$ as
\begin{eqnarray}
 {\cal NLLR} = -log_e~{\cal L}_R = \sum_{k=1}^{k=n_b} n_k \left(\frac{\lambda_k}{n_k} 
+ \frac{1}{2}(\frac{\lambda_k}{n_k})^2
+ \frac{1}{3}(\frac{\lambda_k}{n_k})^3
+ \frac{1}{4}(\frac{\lambda_k}{n_k})^4 \cdots\right)\\
= \sum_{k=1}^{k=n_b} \frac{1}{2}(\frac{\lambda_k^2}{n_k})
+ \frac{1}{3}(\frac{\lambda_k^3}{n_k^2})
+ \frac{1}{4}(\frac{\lambda_k^4}{n_k^3})\cdots
\label{lambchi}
\end{eqnarray}
As $n \rightarrow \infty$, the individual bin contents become normally
distributed about their expected value $T_k$ with variance $\sigma_k^2
= n_k (1-n_k/n)\approx n_k$ for $n_k<<n$. This is true for all cases 
(named the {\it null hypothesis}) where the data and theory fit each other.
Then we can write $\chi_k^2 = \lambda_k/n_k$ and
\begin{equation}
 {\cal NLLR} = \sum_{k=1}^{k=n_b} \frac{1}{2} \chi_k^2 
+ \frac{1}{3}\frac{\lambda_k^3}{\sigma_k^4}
+ \frac{1}{4}\frac{\lambda_k^4}{\sigma_k^6} \cdots
\end{equation}
For large $n$, $\lambda_k\approx \sqrt n_k$ and the higher order terms
 may be neglected yielding
\begin{equation}
{\cal NLLR} \rightarrow \sum_{k=1}^{k=n_b} \frac{1}{2} \chi_k^2 
\: when\:n\rightarrow\infty 
\end{equation}
This is an example of the likelihood ratio theorem~\cite{likeli}.
The expected value of the ${\cal NLLR}$ can then be written
\begin{equation}
 E({\cal NLLR}) = \sum_{k=1}^{k=n_b} \frac{1}{2} E(\chi_k^2)
+ \frac{1}{3}\frac{\mu_3}{\sigma_k^4}
+ \frac{1}{4}\frac{\mu_4}{\sigma_k^6} 
+ \frac{1}{5}\frac{\mu_5}{\sigma_k^8} 
+ \frac{1}{6}\frac{\mu_6}{\sigma_k^{10}} \cdots 
\end{equation}
where $\mu_3,\mu_4,\cdots $ are the $3^{rd}, 4^{th}\cdots$ moments of
the normal distribution about the mean. Since the normal distribution
is symmetric about the mean, all the odd moments ($\mu_3, \mu_5 \cdots
$) are zero. The even moments of the normal distribution (for integer
$l$) are given by the formula
\begin{equation}
 \mu_{2l} = 1.3.5 \cdots(2l-1)\sigma^{2l}
\end{equation}
This yields
\begin{equation}
 E({\cal NLLR}) = \sum_{k=1}^{k=n_b} \frac{1}{2} E(\chi_k^2)
+ \frac{3}{4}\frac{\sigma_k^4}{\sigma_k^6} 
+ \frac{15}{6}\frac{\sigma_k^8}{\sigma_k^{10}} \cdots 
\end{equation}
All the remaining terms tend to zero as $1/n_k(=1/\sigma_k^2)$ as 
$n_k\rightarrow\infty$ leading to
\begin{eqnarray}
 E({\cal NLLR}) = \sum_{k=1}^{k=n_b} \frac{1}{2} E(\chi_k^2) = \frac{n_b}{2} \\
 E ({\cal L}_R )= \exp(-n_b/2) \label{eqbnl}
\end{eqnarray}
The number of degrees of freedom for ${\cal NLLR}$ would be $n_b-1$, 
due to the normalization condition $\sum_k n_k = n$.
\subsection{Normalizing theory and experiment and the problem of
Goodness of fit for the Poisson distribution} 

As we have pointed out,
maximum likelihood fitting only fits the shape of the theoretical
distribution to the experimental data. This is due to the
normalization condition of equation~\ref{thenorm}. However, if we
employ a binomial distribution and define the first bin as 
containing the number of observed events $n$ with theoretical 
expectation of $n_t$ events, and the second bin to
contain the number of unobserved events in $N$ tries, then one can
employ the formula in equation~\ref{lrat2} with $n_b=2$ to obtain the
likelihood ratio.
\begin{equation}
{\cal L}_R= \left(\frac{n_t}{n}\right)^n \left( \frac{N-n_t}{N-n} 
              \right) ^{N-n}
   = \left(\frac{n_t}{n}\right)^n \left( \frac{1-n_t/N}{1-n/N} \right) ^{N-n}
\end{equation}
We now take the Poissonian limit of $N\rightarrow \infty$ with $n_t$ and 
$n$ finite and the above likelihood ratio becomes
\begin{equation}
{\cal L}_R= e^{-(n_t-n)}\left(\frac{n_t}{n}\right)^n 
\label{poiss}
\end{equation}
where we have employed the relations $(N-n)\rightarrow N$ and 
$(1-x/N)^N \rightarrow e^{-x}$ as $N\rightarrow \infty$.

Equation~\ref{poiss} provides the goodness of fit likelihood ratio for
all Poissonian problems where $n_t$ events are expected and $n$ are
observed. We can now multiply this Poissonian ${\cal L}_R$ with
equation~\ref{lrat2} to produce the likelihood ratio for a general
binned likelihood problem where the normalization for theory and
experiment vary.
\begin{equation}
{\cal L}_R= e^{-(n_t-n)}\left(\frac{n_t}{n}\right)^n \:
\prod_{k=1}^{k=n_b} \left ( \frac {T_k}{n_k}\right)^{n_k} 
= e^{-(n_t-n)} \prod_{k=1}^{k=n_b} \left ( \frac {T'_k}{n_k}\right)^{n_k} 
\label{pmult}
\end{equation}
where we have defined $T'_k = n_t T_k/n$ and $\sum T'_k = n_t$.
With this redefinition, we obtain the ${\cal NLLR}$ for the multinomial with
theoretical normalization differing from the experimental one as
\begin{equation}
 {\cal NLLR} = \sum_{k=1}^{k=n_b} T'_k -n_k + n_k\: log_e (\frac{n_k}{T'_k})
\end{equation}
This is same as the ``Poissonian result'' of Baker and Cousins~\cite{baker} 
again derived using very different arguments for the denominator of the 
likelihood ratio.

\subsection{The Gaussian limit of the binomial}
The Poissonian result is useful when $n_t$ and $n$ are relatively
small numbers ($< \approx 25$). When we have larger number of events,
then the Gaussian approximation is more relevant. We have already
shown that (equation~\ref{lambchi}) that in a multinomial, the
negative log likelihood ratio can be approximated
by
\begin{equation}
 {\cal NLLR} = \sum_{k=1}^{k=n_b} 
\frac{1}{2}\left(\frac{\lambda_k^2}{n_k}\right )
\end{equation}
We apply this to the binomial with $n_b=2$, $n_1 = n$, and $n_2= N-n$ and 
$\lambda_1 = -\lambda_2 = n-n_t$. Then
\begin{eqnarray}
 {\cal NLLR} = 
\frac{\lambda_2^2}{2} \left ( \frac{1}{n_1} + \frac{1}{n_2} \right ) =
\frac{\lambda_2^2}{2} \left ( \frac{1}{(1-n/N)(n/N)N} \right )\\ \approx
\frac{\lambda_2^2}{2} \left ( \frac{1}{Npq} \right ) =
\frac{(n-n_t)^2}{2\sigma^2}
\end{eqnarray}
where $p = n_t/N \approx n/N$ is the probability of an 
event appearing in the first bin and $q=1-p$ and $\sigma^2=Npq$ is the 
variance of the bin contents of the first bin. 
We now let $N\rightarrow \infty$, $n\rightarrow \infty$ and $N>>n$. 
In this case, the variance can be approximated by $n$ and we have the 
Gaussian case with  ${\cal NLLR} = (n-n_t)^2/2n)$.
This ${\cal NLLR}$ can be added to the one resulting from the maximum 
likelihood shape fitting to get an overall goodness of fit.

 We must emphasize once again that the method of maximum likelihood
 always fits theoretical shapes to experimental data. We have been
 able to circumvent this restriction by using the device of the
 binomial distribution where the observed events $n$ are in the first
 bin and the total number of events in the distribution $N$ refer to
 the ``number of tries'' and the second bin consists of the $N-n$
 events that failed to appear in the experiment. The binomial
 distribution is special in this regard since once we specify the
 properties of the first bin, the second bin is completely specified
 and anti-correlated with the first bin. The number of tries is
 unknown, but we set it to infinity in two different limits as
 discussed resulting in the Poisson and the Gaussian likelihood
 ratios.

\subsection{To show that $\chi^2$ is also the negative logarithm of a likelihood 
ratio}
The most commonly used method for goodness of fit is
the $\chi^2$ test of Karl Pearson, which is used even when the 
quantities being fitted are not events but measurements with error bars.
We show here that the $\chi^2$ measure is also twice the negative 
logarithm of a Gaussian likelihood ${\it ratio}$ rather than the
 negative logarithm of a Gaussian likelihood, as is the popular misconception.
Consider a binned histogram where the contents in the
$k^{th}$ bin is noted by $c_k$ and the theoretical expectation of this
bin is $s_k$.  The standard error of the observed variable $c_k$ is
 known to be $\sigma_k$. Then, one can write
\begin{equation}
 P(c_k|s_k) = 
 \frac{1}{\sqrt{2\pi} \sigma_k}\exp\left(-\frac{(c_k-s_k)^2}{2\sigma_k^2}\right) = 
 \frac{1}{\sqrt{2\pi} \sigma_k}\exp\left(-\frac{\chi_k^2}{2}\right) 
\end{equation}
This leads to
\begin{equation}
 -\log_e\left( P(c_k|s_k)\right) = \frac{\chi_k^2}{2} + 
                 \log_e(\sqrt{2\pi}\sigma_k)
\end{equation}
From the above expression, people are mistakenly led to conclude that
$\chi^2$ is equivalent to twice the negative log-likelihood. This
ignores the term $\log_e(\sqrt{2\pi} \sigma_k)$ in the above equation,
which varies from bin to bin.  In order to work out the likelihood
ratio, we need to estimate the data density $P(c_k)$ at each
measurement. The data points are distributed as a Gaussian with
standard deviation $\sigma_k$. The best estimate of the mean of the
Gaussian from the data alone is $c_k$. This leads to
\begin{equation}
 P(c_k) = \frac{1}{\sqrt{2\pi}\sigma_k}
\exp\left(-\frac{(c_k-c_k)^2}{2\sigma_k^2}\right) = 
\frac{1}{\sqrt{2\pi}\sigma_k}
\end{equation}
yielding the likelihood ratio
\begin{equation}
{\cal L_R}^k =  \frac{P(c_k|s_k)}{P(c_k)} = 
\exp\left(-\frac{(s_k-c_k)^2}{2\sigma_k^2}\right)= \exp(-\frac{\chi_k^2}{2})
\end{equation}
The overall likelihood ratio is given by
\begin{equation}
{\cal L}_R = \prod_{k=1}^{k=n_b} {\cal L}_R^k
\end{equation}
leading to
\begin{equation}
\chi^2 = 2~log_e \left({\cal L}_R \right) = \sum_{k=1}^{k=n_b} \chi^2_k 
\end{equation}
i.e. $\chi^2$ is equal to twice the negative log-likelihood ratio 
and not the negative log-likelihood!.

\section{Unbinned Goodness of Fit}

Very often the data are not plentiful enough to bin adequately
and it is more
efficient to perform an unbinned likelihood fit. Presently, a goodness
of fit method does not exist for unbinned likelihood fits.  Using the
formalism developed above, we present a solution. After the unbinned
likelihood fit is performed by maximizing the likelihood in
equation~\ref{like} one needs to work out the {\it data likelihood}
$P^{data}(\vec{c_n})$ in order to evaluate the likelihood ratio and
the goodness of fit.  We employ the technique of Probability Density
Estimators $(PDE's)$, also known as Kernel Density
Estimators~\cite{parzen} $(KDE's)$ to do this. The $pdf$ $P^{data}(c)$
is approximated by
\begin{equation}
  P^{data}(c)\approx PDE(c) = \frac{1}{n}\sum_{i=1}^{i=n} {\cal G}(c-c_i)
\label{pde}
\end{equation}
where a Kernel function ${\cal G}(c-c_i)$ is centered around each data 
point $c_i$, is so defined that it normalizes to unity. The choice of the 
Kernel function can vary depending on the problem. A popular kernel is the 
Gaussian defined in the multi-dimensional case as
\begin{equation}
 {\cal G}(c) = \frac{1}{(\sqrt{2\pi}h)^d\sqrt(det(E))}
exp(\frac{-H^{\alpha\beta}c^\alpha c^\beta}{2h^2})
\end{equation}
where $E$ is the error matrix of the data defined as
\begin{equation}
 E^{\alpha,\beta} = <c^\alpha c^\beta > -<c^\alpha><c^\beta>
\end{equation}
and the $<>$ implies average over the $n$ events, and $d$ is the
 number of dimensions. The Hessian matrix $H$ 
is defined as the inverse of $E$ and 
the repeated indices imply summing over.  The parameter $h$ is a 
``smoothing parameter'',  which has\cite{hoptim} a 
suggested optimal value $h \propto n^{-1/(d+4)}$, that 
satisfies the asymptotic condition 
\begin{equation}
  {\cal G}_\infty(c-c_i)\equiv\lim_{n \rightarrow \infty} {\cal G}(c-c_i) = \delta (c-c_i)
\end{equation}
The parameter $h$ will depend on the local number density and will have to be 
adjusted as a function of the local density to obtain good representation of 
the data by the $PDE$.
Our proposal for the goodness of fit in unbinned likelihood fits is thus 
the likelihood ratio
\begin{equation}
{\cal L_R} = \frac{P(\vec{c_n}|s)}{P^{data}(\vec{c_n})} \approx 
\frac{P(\vec{c_n}|s)}{P^{PDE}(\vec{c_n})}
\end{equation}
evaluated at the maximum likelihood point $s^*$.
\subsection{An illustrative example}
\begin{figure}[t]
\centerline{\includegraphics[width=\textwidth]{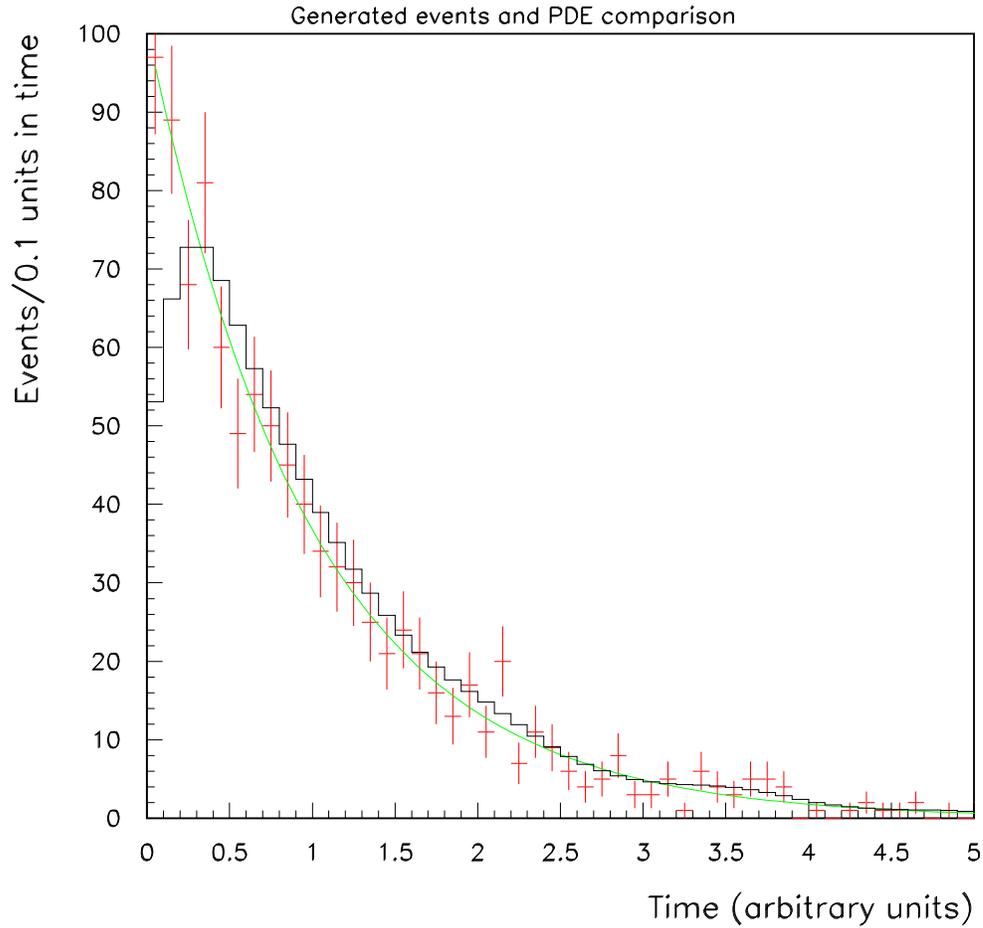}}
\caption
{Figure shows the histogram (with errors) of generated events. 
Superimposed is the theoretical
curve $P(c|s)$  and the $PDE$ estimator (solid) histogram with no errors.
\label{genev}}
\end{figure}

We consider a simple one-dimensional case where the data is an exponential distribution, say decay times of a radioactive isotope. 
The theoretical prediction is given by
\begin{equation}
P(c|s) = \frac{1}{s}\exp(-\frac{c}{s})
\end{equation}
We have chosen an exponential with $s=1.0$ for this example.
The Gaussian Kernel for the $PDE$ would be given by
\begin{equation}
{\cal G}(c) = \frac{1}{(\sqrt{2\pi} \sigma h)}
\exp (-\frac{c^2}{2\sigma^2h^2})
\end{equation}
where the variance $\sigma$ of the exponential is numerically equal to $s$.
To begin with, we chose a constant value for the smoothing parameter, 
which for 1000 events generated is calculated to be 0.125.
Figure~\ref{genev} shows the generated events, the theoretical curve
$P(c|s)$ and the $PDE$ curve $P(c)$ normalized to the number of
events. The $PDE$ fails to reproduce the data near the origin due to the 
boundary effect, whereby the 
Gaussian probabilities for events close to the origin 
spill over to negative values of $c$. 
This lost probability would be compensated by events on the
exponential distribution with negative $c$ if they existed. In our
case, this presents a drawback for the $PDE$ method, which we will
remedy later in the paper using $PDE$ definitions on the  hypercube
and periodic boundary conditions. For the time being, we will confine our 
example to values of $c > 1.0$ to avoid the boundary effect. 

 In order to test the goodness of fit 
capabilities of the likelihood ratio ${\cal L_R}$, we superimpose a 
Gaussian on the exponential and try and fit the data by a simple exponential.
Figure~\ref{genev1} shows the ``data'' with 1000 events generated as
an exponential in the fiducial range $1.0 <c< 5.0$. Superimposed on it 
is a Gaussian of 500 events. More events in the exponential are generated 
in the interval $0.0 <c < 1.0$ to avoid the boundary effect at the 
fiducial boundary at c=1.0.
Since the number
density varies significantly, we have had to introduce a method of
iteratively determining the smoothing factor as a function of $c$. 

\subsection{Iterative Determination of the Smoothing Factor}

The expression $h\approx n^{-1/(d+4)}$ clearly is meant to give a
smoothing factor that decreases slowly with increased statistics
$n$. It is expected to be true on average over the whole 
distribution. However, the exponential distribution under consideration
has event densities that vary by orders of magnitude as a function of
the time variable $c$. In order to obtain a function $h(c)$ that takes
into account this variation, we first work out a $PDE$ with
constant $h$ and then use the number densities obtained thus~\cite{power} 
to obtain $h(c)$ as per the equation
\begin{equation}
  h(c) =   \left(\frac{n\:PDE(c)}{(c_2-c_1)}\right)^{-0.6}
\end{equation}

The equation is motivated by the consideration that a uniform
distribution of events between $c_1$ and $c_2$ has a $pdf= 1/(c_2-c_1)$
whereas the real $pdf$ is approximated by $PDE$. The function
$h(c)$ thus obtained is used to work out a better $PDE(c)$. This process 
is iterated three times to give the best smoothing function.

\begin{figure}[tbh!]
\centerline{\includegraphics[width=4.0in]{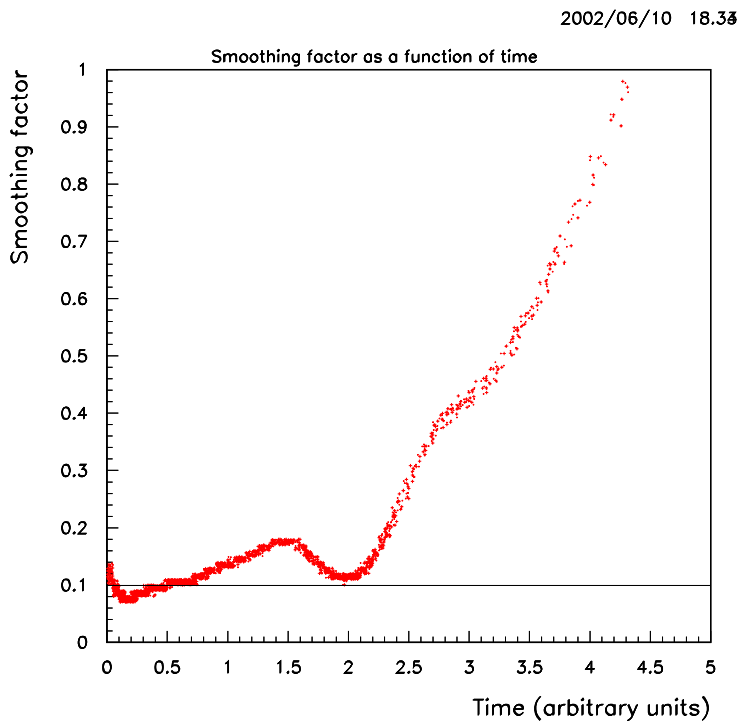}}
\caption[smoothing factor as a function of $c$]
{The variation of $h$ as a function of $c$ for the example shown in 
Figure~\ref{genev1}. The variation of the smoothing parameter is obtained 
iteratively as explained in the text. The flat curve is a smoothing 
factor resulting from the formula $h\approx 0.5n^{-1/(d+4)}$.
\label{hc}}
\end{figure}

We generate $n$=1000 events in the fiducial interval. If now we
were to superimpose a Gaussian with 500 events centered at $c$=2.0 and
width=0.2 on the data, the $PDE$ estimator will follow the data as
shown in Figure~\ref{genev1}. This shows that the $PDE$ estimator we have 
is adequate to reproduce the data, once the smoothing parameter is 
made to vary with the number density appropriately.
The smoothing function $h(c)$ for the events in Figure~\ref{genev1}
is shown in Figure~\ref{hc}.  It can be seen that the value of $h$
increases for regions of low statistics and decreases for regions of
high statistics. Superimposed is the constant smoothing factor
obtained by the equation $h \approx 0.5 n^{-1/(d+4)}= 0.5n^{-0.2}$,
with $n$ being the total number of events generated, including those
outside the fiducial volume.
\subsection{An Empirical Measure of the Goodness of Fit}

\begin{figure}[tbh!]
\centerline{\includegraphics[width=4.0in]{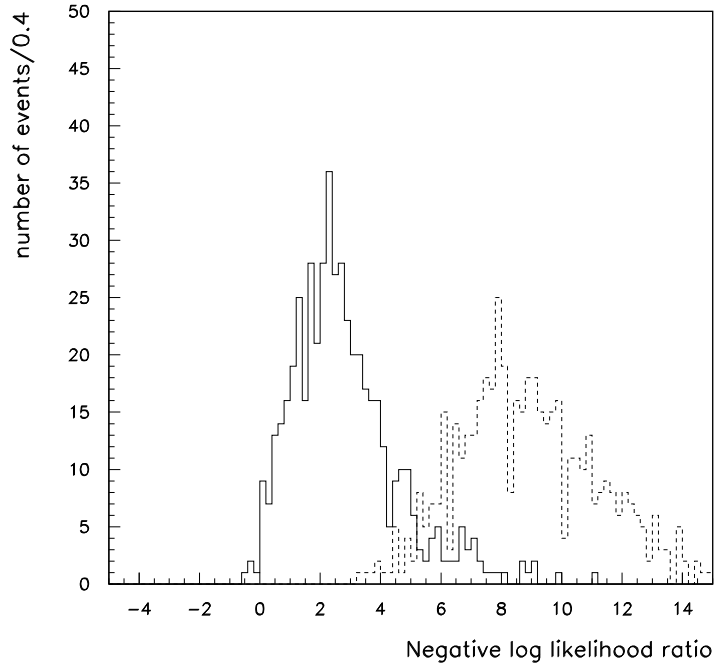}}
\caption[Distribution of fitted negative log-likelihood ratios]
{The solid curve shows the distribution of the negative log likelihood
ratio ${\cal NLLR}$ at the maximum likelihood point for 500
distributions, using the iterative smoothing function mechanism. The
dashed curve shows the corresponding distribution in the case of a
constant smoothing function.
\label{fitlike}}
\end{figure}

The negative log-likelihood ratio ${\cal NLLR} \equiv -{log_e {\cal
L_R}}$ at the maximum likelihood point now provides a measure of the
goodness of fit. In order to use this effectively, one needs an analytic theory
of the sampling distribution of this ratio. This is difficult to
arrive at, since this distribution is sensitive to the smoothing
function used. If adequate smoothing is absent in the tail of the
exponential, larger and broader sampling distributions of ${\cal
NLLR}$ will result. 
One can however determine the distribution of ${\cal NLLR}$
empirically, by generating the events distributed according to the
theoretical model many times and determining ${\cal NLLR}$ at the
maximum likelihood point for each such distribution. The solid
histogram in figure~\ref{fitlike} shows the distribution of ${\cal
NLLR}$ for 500 such fits.  This has a mean of 2.8 and an $rms$ of 1.8. The
dotted histogram shows the corresponding value of ${\cal NLLR}$ for
the constant value of smoothing factor shown in figure~\ref{hc}. This
distribution is clearly broader ($rms$=2.63) with a higher mean(=9.1) 
and thus has less discrimination power in judging the goodness of fit 
than the solid curve.

\begin{figure}[tbh!]
\centerline{\includegraphics[width=\textwidth]{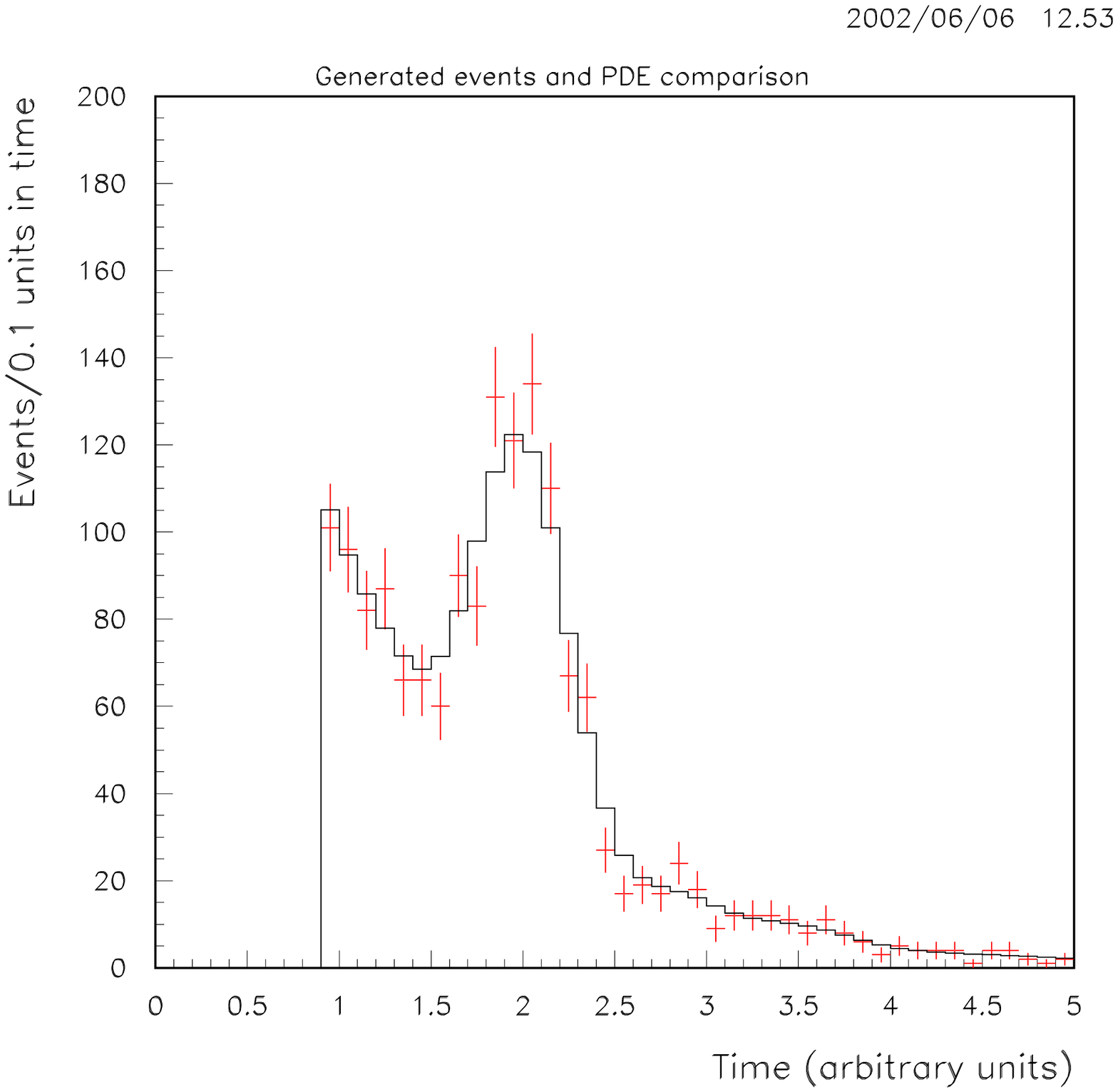}}
\caption[generated events]
{Figure shows the histogram (with errors) of 1000 events in the 
fiducial interval $1.0<c<5.0$ generated as
an exponential with decay constant $s$=1.0 with a superimposed
Gaussian of 500 events centered at $c$=2.0 and width=0.2.  
The $PDE$ estimator is
the (solid) histogram with no errors. 
\label{genev1}}
\end{figure}

With this modification in the $PDE$, one
gets a good description of the behavior of the data by the $PDE$ as
shown in Figure~\ref{genev1}.
We now vary the number of events in the Gaussian and obtain the value
of the negative log likelihood ratio ${\cal NLLR}$ as a function of
the strength of the Gaussian.  Table~\ref{tab1} summarizes the
results. The number of standard deviations the unbinned likelihood fit is from
what is expected is determined empirically by plotting the value of
${\cal NLLR}$ for a large number of fits where no Gaussian is
superimposed (i.e. the null hypothesis) and determining the mean and
$RMS$ of this distribution and using these to estimate the number of
$\sigma$'s the observed ${\cal NLLR}$ is from the null
case. Table~\ref{tab1} also gives the results of a binned fit on the
same ``data''. It can be seen that the unbinned fit gives a $3\sigma$
discrimination when the number of Gaussian events is 85, where as the
binned fit gives a $\chi^2/ndf$ of 42/39 for the same case. 
\begin{table}[bht!]
\caption{
Goodness of fit results from unbinned likelihood and binned likelihood fits for
various data samples. The negative values for the number of 
standard deviations in some of the examples is due to statistical fluctuation.
\label{tab1}}
\centering\leavevmode
\begin{tabular}{|c|c|c|c|}
\hline
Number of & Unbinned fit & Unbinned fit& Binned fit $\chi^2$ \\
Gaussian events & ${\cal NLLR}$ & $N\sigma$ & 39 d.o.f.\\
\hline
500 & 189. & 103 & 304 \\
250 & 58.6 & 31 & 125 \\
100 & 11.6 & 4.9& 48 \\
85 & 8.2 & 3.0 & 42 \\
75 & 6.3 & 1.9 & 38 \\
50 & 2.55 & -0.14 &30 \\
0  & 0.44 & -1.33 & 24 \\
\hline
\end{tabular}
\end{table}

\begin{figure}[tbh!]
\centerline{\includegraphics[width=\textwidth]{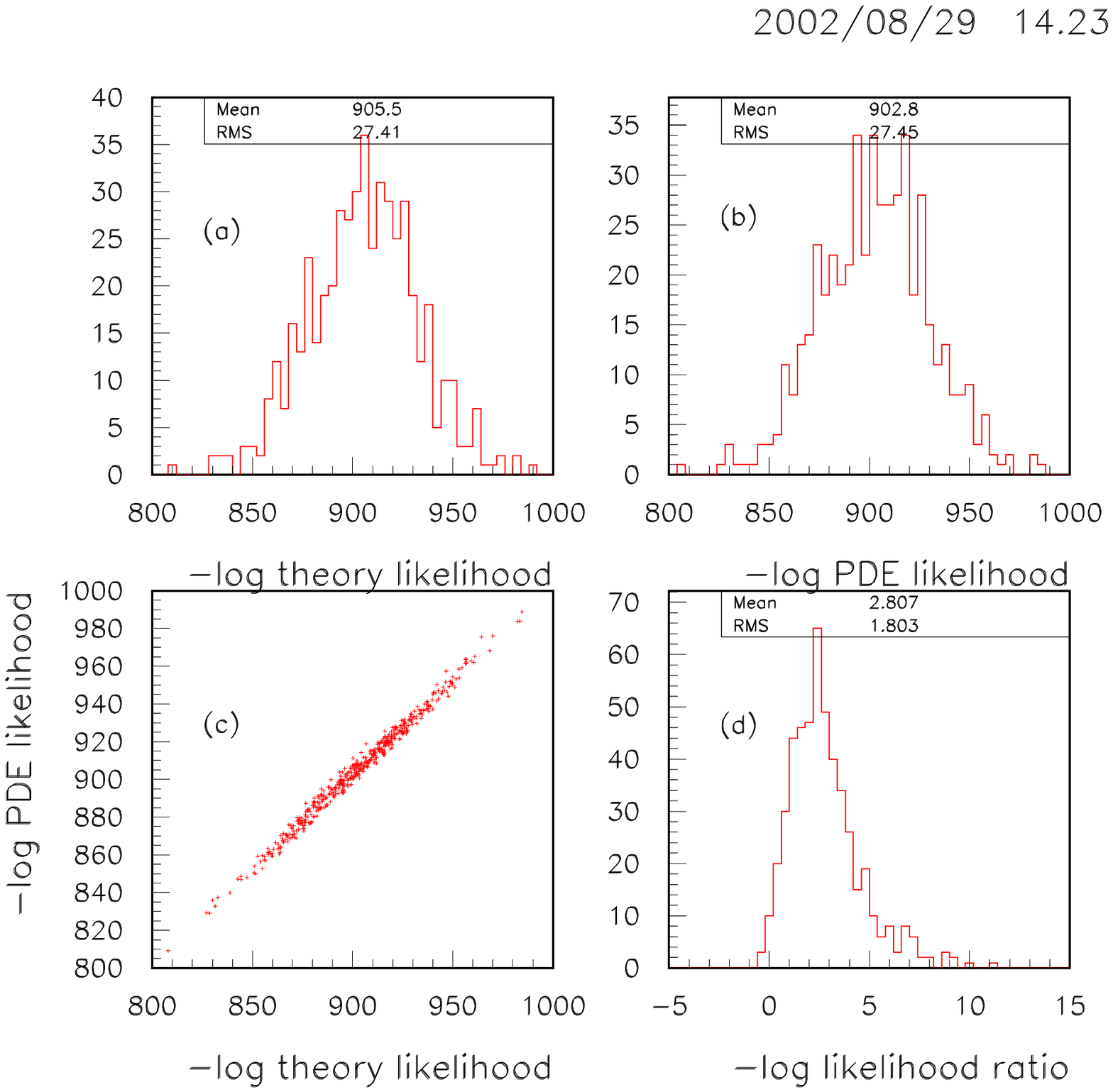}}
\caption
{(a) shows the distribution of the negative log-likelihood -$log_e
(P(\vec{c_n}|s))$ for an ensemble of experiments where data and experiment 
are expected to fit. (b) Shows the negative log $PDE$ likelihood -$log_e
(P(\vec{c_n}))$ for the same data (c) Shows the correlation between the
two and (d) Shows the negative log-likelihood ratio ${\cal NLLR}$ that
is obtained by subtracting (b) from (a) on an event by event basis.
\label{compar}}
\end{figure}

 Figure~\ref{compar} shows the variation of -log $P(\vec{c_n}|s)$ and
-log $P^{PDE}(\vec{c_n})$ for an ensemble of 500 experiments each with
the number of events $n=1000$ in the exponential and no events in the
Gaussian (null hypothesis). One notes that -log $P(\vec{c_n}|s)$ and
-log $P^{PDE}(\vec{c_n})$ are correlated with each other and the difference 
between the two (-log ${\cal NLLR}$) is a much narrower distribution 
than either and provides the goodness of fit discrimination.
\subsection{Improving the $PDE$}
The $PDE$ technique we have used so far suffers from two drawbacks;
firstly, the smoothing parameter has to be iteratively adjusted
significantly over the full range of the variable $c$, since the
distribution $P(c|s)$ changes significantly over that range; and
secondly, there are boundary effects at $c$=0 as shown in
figure~\ref{genev}.  Both these flaws are remedied if we define the
$PDE$ in hypercube space.  After we find the maximum likelihood point
$s^*$, for which the $PDE$ is not needed, we transform the variable
$c\rightarrow c'$, such that the distribution $P(c'|s^*)$ is flat and
$0<c'<1$. The hypercube transformation can be made even if $c$ is 
multi-dimensional by initially going to a set of variables that are
 uncorrelated and then making the hypercube transformation. 
The transformation can be such that any interval in $c$ space maps on
to the interval $(0,1)$ in hypercube space.
\subsection{Periodic Boundary Conditions}
We solve the boundary
problem by imposing periodicity in the hypercube.  In the one
dimensional case, we imagine three ``hypercubes'', each identical to the
other on the real axis in the intervals $(-1,0)$, $(0,1)$ and $(1,2)$. The
hypercube of interest is the one in the interval $(0,1)$.  When the
probability from an event kernel leaks outside the boundary $(0,1)$, we
continue the kernel to the next hypercube. Since the hypercubes are
identical, this implies the kernel re-appearing in the middle
hypercube but from the opposite boundary. Put mathematically, the kernel 
is defined such that
\begin{eqnarray}
 {\cal G}(c'-c'_i) ={\cal G}(c'-c'_i-1);\: c'>1 \\
 {\cal G}(c'-c'_i) ={\cal G}(c'-c'_i+1);\: c'<0 
\end{eqnarray}

Although a Gaussian Kernel will work on the hypercube, 
the natural kernel to use considering the shape of the distribution in 
hypercube space (it is flat for a good fit), 
would be  the ``boxcar function''  ${\cal G}(c')$.
\begin{eqnarray}
 {\cal G}(c') = \frac{1}{h} ;\: |c'|<\frac{h}{2} \\
 {\cal G}(c') = 0 ;\: |c'|>\frac{h}{2} 
\end{eqnarray} 
This kernel would be subject to the periodic boundary conditions given above, 
which further ensure that every configuration in hypercube space is treated
 exactly as every other configuration irrespective of its co-ordinate.
The parameter $h$ is a smoothing parameter which needs to be chosen 
with some care. However, since the theory distribution is flat in 
hypercube space, the smoothing parameter may not need to be iteratively 
determined over hypercube space to the extent that 
data distribution is similar to 
the theory distribution. Even if iteration is used, the variation in $h$ 
in hypercube space is likely to be much smaller.
\begin{figure}[tbh!]
\centerline{\includegraphics[width=\textwidth]{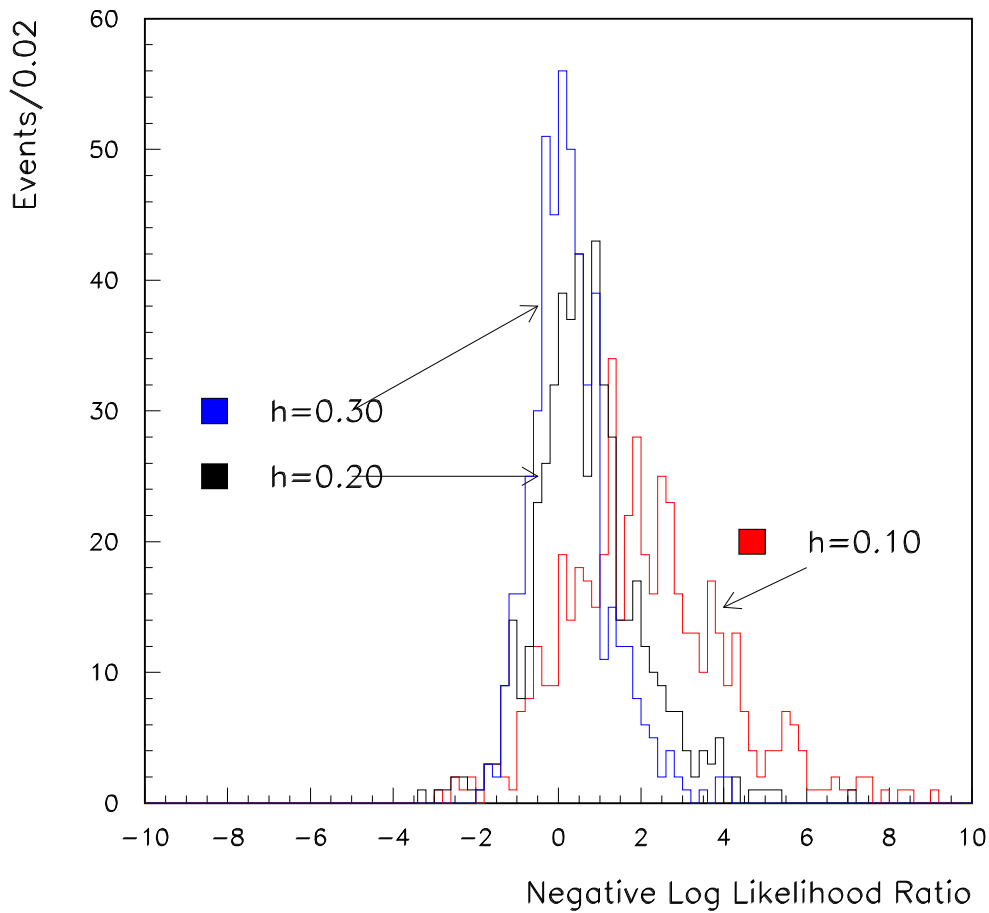}}
\caption
{The distribution of the negative log likelihood ratio ${\cal NLLR}$ for 
the null hypothesis for an ensemble of 500 experiments each with 1000 events, 
as a function of the smoothing factor  $h$=0.1, 0.2 and 0.3
\label{hyperbin}}
\end{figure}

\begin{figure}[tbh!]
\centerline{\includegraphics[width=\textwidth]{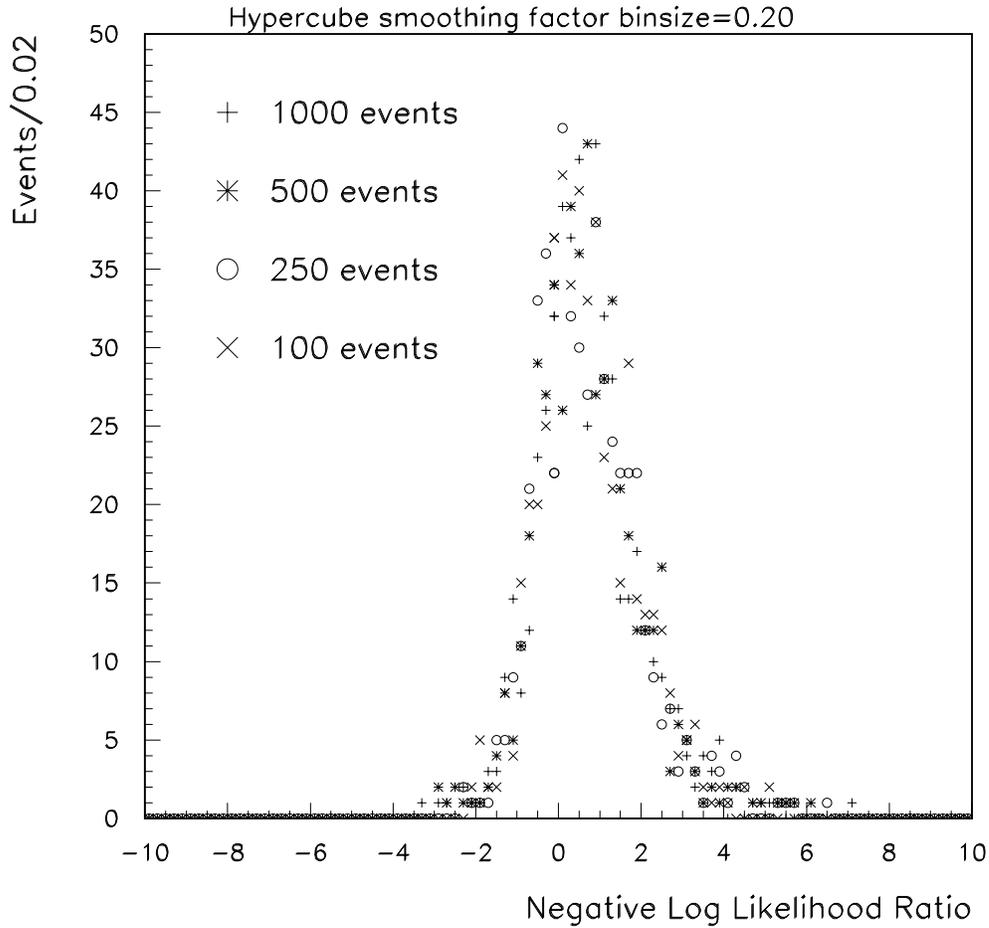}}
\caption
{The distribution of the negative log likelihood ratio ${\cal NLLR}$ for 
the null hypothesis for an ensemble of 500 experiments each with
the smoothing factor  $h$=0.2, as a function of the number of events 
\label{hyperev}}
\end{figure}

 Figure~\ref{hyperbin} shows the distribution of the ${\cal NLLR}$ for the 
null hypothesis for an ensemble of 500 experiments each with 1000 events as a 
function of the smoothing factor $h$. It can be seen that the distribution 
narrows considerably as the smoothing factor increases. We choose an operating
 value of 0.2 for $h$ and study the dependence of the ${\cal NLLR}$ as a 
function of the number of events ranging from 100 to 1000 events, as shown in 
figure~\ref{hyperev}. The 
dependence on the number of events is seen to be weak, indicating 
good behavior. The $PDE$ thus arrived computed with $h$=0.2 can be transformed
 from the hypercube space to $c$ space and will reproduce data smoothly and 
with no edge effects. We note that it is also easier to arrive at an 
analytic theory of ${\cal NLLR}$ with the choice of this simple kernel.
\section{The distribution of the goodness of fit variable}

Of all the goodness of fit variables we have studied above, for both
binned and unbinned likelihood fits, the $\chi^2$ variable is the most
studied and has an analytic theory associated with its
distribution. This is used to set a $p$-value for the goodness of fit,
defined as the probability to exceed the observed value $\chi^2$ based
on its analytic distribution. In the absence of an analytic theory, it
is possible to use Monte Carlo methods to obtain the distribution of
the goodness of fit variable for the hypothesis being tested and to
numerically obtain the $p$-value.
 
\section{Calculation of fitted errors}

After the fitting is done and the goodness of fit is evaluated, one
needs to work out the errors on the fitted quantities. One needs to
calculate the probability density $P(s|\vec{c_n})$, 
which carries information not 
only about the maximum likelihood point $s^*$, from a single experiment,
 but how  such a measurement is likely to fluctuate if we 
repeat the experiment. This problem is known as the ``problem of inverse
 probabilities'' in statistical literature and is solved by the use of
 Bayes' theorem. Since Bayes' theorem is central to the arguments that 
follow, we give a simple derivation of it here.
\subsection{Derivation of Bayes' theorem equations}

\begin{figure}[tbh]
\centerline{\includegraphics[width=4.0in]{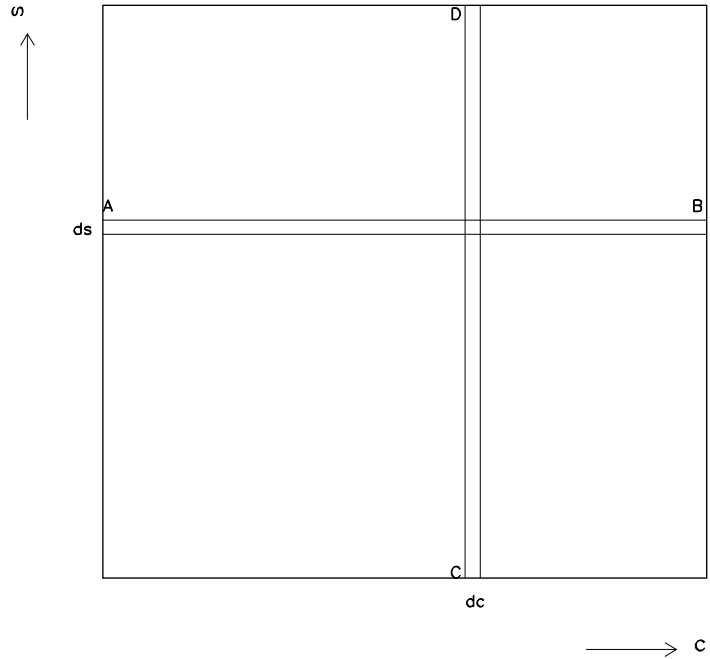}}
\caption[]
{Joint probability distribution in the variables $s$, $c$. Conditional
probabilities are computed along the slices AB( $s$=constant) and
CD($c$= constant).  }
\label{joint}
\end{figure}

Consider a joint probability distribution $P(s,c)$ in variables $s,c$. For the
sake of simplicity, we will take both $s$ and $c$ to be
one-dimensional. The arguments being made are general enough to easily
change them into multi-dimensional variables.
Figure~\ref{joint} shows geometrically the two dimensional space of
$s$ and $c$. We plot $s$ as the ordinate and $c$ as the abscissa. At
this stage $s$ and $c$ are two general variables. Then,
\begin{equation}
 \int \int P(s,c) ds dc =1 
\label{norm}
\end{equation}
We define the single variable  probabilities $P(c)$ and $P(s)$ as
\begin{eqnarray}
P(c) = \int P(s,c) ds \label{eq1}\\
P(s) = \int P(s,c) dc \label{eq2}
\end{eqnarray}
$P(c)$ is the probability density of $c$ irrespective of the value of
$s$ and $P(s)$ is the probability density of $s$ irrespective of the
value of $c$. It follows from equation~\ref{norm} that
\begin{equation}
 \int P(s) ds =1 
\end{equation}
and
\begin{equation}
 \int P(c) dc =1 
\end{equation}
We define a conditional probability $P(c|s)$ as the probability of
observing $c$ given $s$. It is thus, the joint probability $P(s,c)$
along the slice AB ($s$=constant) in figure~\ref{joint}, appropriately
normalized to unity. $i.e$,
\begin{equation}
 P(c|s) = \frac{P(s,c)}{\int P(s,c)dc}
\end{equation}
where the denominator in the above equation ensures that $\int P(c|s) dc =1$.
Therefore, (using equation~\ref{eq2})
\begin{equation}
 P(c|s) = \frac{P(s,c)}{P(s)}
\label{ps1}
\end{equation}
By symmetrical arguments (integrations along the slice CD), we show that the 
conditional probability $P(s|c)$ is given by
\begin{equation}
 P(s|c) = \frac{P(s,c)}{P(c)}
\label{ps2}
\end{equation}
leading to the joint probability equation
\begin{equation}
 P(s,c) = P(c|s)P(s) = P(s|c)P(c)
\label{jp}
\end{equation}
which is sometimes written in a more familiar form known as Bayes'
theorem~\cite{bayes} as
\begin{equation}
 P(s|c) = \frac{P(c|s)P(s)}{P(c)}
\end{equation}
By substituting the
expression for  $P(s,c)$ in equation~\ref{ps1} in
equation~\ref{eq1} we get the equation
\begin{equation}
  P(c) = \int P(c|s)P(s) ds
\label{eqpc}
\end{equation}
and by substituting the expression for $P(s,c)$ in equation~\ref{ps2}
in equation~\ref{eq2} we get the equation
\begin{equation}
  P(s) = \int P(s|c)P(c) dc
\label{eqps}
\end{equation}
These complete the Bayes' theorem equations.
Note also that the joint
probability equation~\ref{jp} can be written in a form a likelihood
ratio ${\cal L}_R$
\begin{equation}
 {\cal L}_R = \frac{P(s|c)}{P(s)} = \frac{P(c|s)}{P(c)}
\label{lr}
\end{equation}
The quantity ${\cal L}_R$ equation~\ref{lr} is invariant under change of
variables $c\rightarrow c'$ and $s\rightarrow s'$, since the Jacobian
of the transformation $|\frac{\partial c'}{\partial c}|$ divides out
in the numerator and the denominator for the right hand side of the
equation~\ref{lr} for the ratio of probability densities in
$\frac{P(c|s)}{P(c)}$.  Similarly the ratio is invariant under the
transformation variable $s$ in the LHS of the equation. These
invariances are essential in the use of the ratio ${\cal L}_R$ as a
goodness-of-fit variable.

We can then extend the derivation given above to derive Bayes' theorem 
equations for the dataset $\vec{c_n}$. 
\begin{eqnarray}
 P(s,\vec{c_n}) = P(\vec{c_n}|s)P(s) = P(s|\vec{c_n})P(\vec{c_n}) \label{eqjnt}\\
 P(\vec{c_n}) = \int P(\vec{c_n}|s) P(s) ds\label{eqpcn}\\
  P(s) = \int P(s|\vec{c_n})P(\vec{c_n}) d\vec{c_n}
\label{eqpn}
\end{eqnarray}

Let us  note that Bayes' theorem as derived above is a theorem in
mathematics that applies to any integrable function of two variables
$P'(s,c)$ for which the normalized function $P(s,c)
=\frac{P'(s,c)}{\int\int P'(s,c)ds dc}$ can be constructed. The
``probabilities'' $P(s)$ and $P(c)$ are projections of such a function
and the ``conditional probabilities'' $P(c|s)$ and $P(s|c)$ are
normalized slices of such a function. How we identify the slices and
projections to correspond to probability theory is up to us. Once the
joint probability function $P(s,c)$ can be specified, the inverse
probability problem can be solved, since the inverse probability is
given by the slice $P(s|c)$ and the theoretical likelihood is given by
the slice $P(c|s)$.
In order to specify the joint probability, the Bayesians specify the
theoretical likelihood $P(\vec{c_n}|s)$ and the projection on the parameter
axis $P(s)$ which they term the Bayesian prior. We now describe the
Bayesian paradigm and the difficulties associated with it.
\subsection{The Bayesian Paradigm}
The theoretical likelihood $P(\vec{c_n}|s)$ is specified by the model in
question. In order to specify the joint probability, Bayesians supply
the function $P(s)$, which they term the Bayesian prior. The joint
probability is then given by

\begin{equation}
P(s,\vec{c_n}) = P(\vec{c_n}|s)P(s) = P(s|\vec{c_n})P(\vec{c_n})
\end{equation}
and the inverse probability is obtained by
\begin{equation}
P(s|\vec{c_n}) = =\frac{P(\vec{c_n}|s)P(s)}{P(\vec{c_n})}= \frac{P(\vec{c_n}|s)P(s)}{\int P(\vec{c_n}|s)P(s)ds}
\end{equation}
The Bayesian prior $P(s)$ is an unknown function that according to the
Bayesians encapsulates the prior knowledge the experimenter has on the
true value of the parameter. There exists many different methods of
estimating the unknown prior. We briefly describe them here.
\subsubsection{Objective Bayesianism}
In this sub-branch of Bayesianism, the prior is assumed to be known
within some limits that the user supplies and the distribution is
assumed to be flat within those limits. This approach is championed by
statisticians such as Jaynes~\cite{jaynes}. If one specifies a flat
prior in a variable $s$, then it is clearly not a flat prior in a
transformed variable $s'=s'(s)$. The maximum likelihood analysis can
be carried out in any function $s'(s)$ and the maximum likelihood
point $s^*$ would be the same under such transformations, i.e.
\begin{equation}
 s^{*'}=s^{'}(s^{*})
\end{equation}
However, if a flat Bayesian prior is assumed in one variable, it would
not be flat in any function of that variable. The results would depend
on the prior assumed. This is a serious objection to this method.
\subsubsection{Subjective Bayesianism}
This approach is championed by statisticians such as de~Finetti~\cite{defi} 
where the experimenter specifies the prior
based on ``subjective criteria'' based on his past experience and
knowledge of the parameter $s$. If more than one experimenter is
involved, then more than one prior can be used and more than one
posterior density results.
\subsubsection{Hierarchical Bayesianism}
This sub-branch of Bayesianism attempts to parametrize the
prior~\cite{hier} in terms of more unknown parameters each of which
have their own priors, forming a hierarchy.
\subsubsection{Empirical Bayesianism}
Empirical Bayesianism~\cite{empir} attempts to stem the infinite
hierarchy of priors implied by hierarchical Bayesianism by attempting
to determine some of the parameters associated with the priors from
the data.

Note all Bayesians~\cite{defi1} treat the projection $P(\vec{c_n})$ as 
an uninteresting constant of normalization, theoretically obtained 
by the equation
\begin{equation}
 P(\vec{c_n}) = \int P(\vec{c_n})|s)P(s)ds
\end{equation}
whose right hand side consists of an integral over the Bayesian prior and 
the theoretical likelihood.
However, having solved the goodness of fit problem, we have
demonstrated that the data likelihood $P^{data}(\vec{c_n})$ carries
with it vital information relevant for goodness of fit. It may be
thought of as the $pdf$ of the $n-object$ $\vec{c_n}$ and must be
empirically determined from experimental data. If we use this function
from the data as a projection of the joint probability 
and the theoretical likelihood
$P(\vec{c_n}|s)$ as a slice, then we can invert the probability to obtain
$P(s|\vec{c_n})$ without the use of a Bayesian prior. What results is
a new paradigm in statistics where we have to re-define some concepts
to be consistent with this interpretation.

\subsection{The New Paradigm}
We note that if we identify the projection $P(\vec{c_n})$ of the joint
probability $P(s,\vec{c_n})$ as the data-likelihood, which we denote
$P^{data}(\vec{c_n})$, then the projection on the $s$ axis depends on
$n$ and is thus incompatible with being a Bayesian prior that is
independent of $n$.

To show this, let us note that the inverse probability
$P(s|\vec{c_n})\rightarrow \delta(s-s_T)$ as $n\rightarrow\infty$
where $s_T$ is the true value of s. This is a result of the central
limit theorem of statistics and contains the assumption that the
experiment is unbiased.
Then, using equation~\ref{eqpn} and $P^{data}(\vec{c_n})$ for one of the 
projections, one obtains
\begin{equation}
P(s) = \int P(s|\vec{c_n})  P^{data}(\vec{c_n}) d\vec{c_n} 
\end{equation}
But, $ P^{data}(\vec{c_n}) d\vec{c_n}$ represents the probability of
obtaining the dataset $\vec{c_n}$ in the neighborhood of the dataset
$\vec{c_n}$. If one were to repeat the experiment $N$ times, 
thus obtaining an ensemble of datasets, then $ P^{data}(\vec{c_n})
d\vec{c_n}=\frac{dN}{N}$ for $N\rightarrow \infty$.  Then,

\begin{equation}
P(s) = \int P(s|\vec{c_n}) \frac{d N}{N} =
\frac{1}{N}\sum_{k=1}^{k=N}P_k(s|\vec{c_n})=<P(s|\vec{c_n})>
\end{equation}
where  $k$
denotes the ensemble member and
the symbols $<>$ represent average over the ensemble of the
function.  However, since $P_k(s|\vec{c_n})\rightarrow \delta (s-s_T)$,
as $n\rightarrow \infty$, we would expect $P(s)\rightarrow
\delta(s-s_T)$ in this limit. i.e. $P(s)$ cannot find interpretation as an 
$n$-independent Bayesian prior. 
We note however, that if one were to plot the probability distribution
of the maximum likelihood value $s^*$ of each member of the ensemble,
then such a distribution would have the desired $n$ dependence, 
becoming narrower for larger n. We
thus build our theory by identifying the projection on the parameter
axis as the probability distribution of $s^*$, which we denote by
${\cal P}_n(s^*)$, explicitly indicating the $n$ dependence. 
Then the joint probability distribution $P(s^*,\vec{c_n})$ is given by
\begin{equation}
P(s^*,\vec{c_n}) = P(\vec{c_n}|s^*){\cal P}_n(s^*)
\label{jnew}
\end{equation}
Each member $k$ of the ensemble has a maximum likelihood value
$s^*_k$. The probability distribution of this quantity over an
infinite ensemble is defined to be ${\cal P}_n(s^*)$. This definition
is similar in spirit to the ``fiducial probability'' of
R.~A.~Fisher~\cite{fiduc}.
We are now able to specify the joint probability $P(s^*,\vec{c_n})$ as per 
equation~\ref{jnew}. Then by Bayes' theorem, we can also write
\begin{equation}
P(s^*,\vec{c_n}) = P(s^*|\vec{c_n})P^{data}(\vec{c_n})=
P(\vec{c_n}|s^*){\cal P}_n(s^*)
\end{equation}
where this equation is the definition of the inverse probability $
P(s^*|\vec{c_n})$. This implies that from the $k^{th}$ element of the ensemble,
consisting of a single dataset
$\vec{c_n}$, not only is the maximum likelihood value $s^*_k$
available, but also information on the distribution of $s^*$ from other
similar datasets on the ensemble. It is the availability of this
information that permits the estimation of errors based on one
dataset. This then leads to the solution of the inverse probability on
the ensemble by the usual Bayes' theorem equation.
\begin{equation}
P(s^*|\vec{c_n}) = \frac{P(\vec{c_n}|s^*){\cal P}_n(s^*)}{P^{data}(\vec{c_n})}=
\frac{P(\vec{c_n}|s^*){\cal P}_n(s^*)}{\int P(\vec{c_n}|s^{*'}){\cal P}_n(s^{*'})ds^{*'}}
\label{invnew}
\end{equation}
Equation~\ref{invnew} shows us how to obtain the inverse probability
$P(s^*|\vec{c_n})$ once we have the ensemble and hence is not much use
to us, since it requires an infinite number of similar experiments on
the ensemble. Our problem is to obtain the inverse probability given a
single member of the ensemble. Before we proceed to solve this
problem, let us note that on the ensemble, Bayes' theorem is expressed
in the following elegant set of equations.
\begin{eqnarray}
{\cal P}_n(s^*) = \int
P(s^*|\vec{c_n})P^{data}({\vec{c_n})d\vec{c_n}=<P(s^*|\vec{c_n})}>\label{lpn}\\
{\cal P}(\vec{c_n}) = \int P(\vec{c_n}|s^*){\cal
P}_n(s^*)ds^*=<P(\vec{c_n}|s^*)>\label{lpc}\\
{\cal L}_R^k(s^*) = \frac{P_k(s^*|\vec{c_n})}{<P(s^*|\vec{c_n})>} 
= \frac{P_k(\vec{c_n})|s^*)}{<P(\vec{c_n})|s^*)>}\label{lrnew} 
\end{eqnarray}
In the above set of equations, we have used the symbol ${\cal
P}(\vec{c_n})$ to denote the data $pdf$ determined on an infinite
ensemble to distinguish it 
from $P_k^{data}(\vec{c_n})$ which is the data pdf determined
from a single member of the ensemble. The former benefits from the
statistics present in the infinite ensemble. Equation~\ref{lrnew}
gives the the likelihood ratio on the ensemble of each member of the
ensemble that may be used for goodness of fit once the ensemble is
known. Note that there is no Bayesian prior used anywhere in the above
set of equations.
\subsubsection{The true value of the parameter $s$}

The true value $s_T$ of the parameter $s$ is defined to be that value
of $s^*$ at which the maximum of the $pdf$ ${\cal P}_n(s^*)$ occurs. Let
us remember that ${\cal P}_n(s^*)$ has an infinite number of similar
datasets $\vec{c_n}$ contributing to it and hence this is just a
statement of the experiments being unbiased. The true value is a number. 
It does not possess a distribution.

\subsubsection{The unknowability of ${\cal P}_n(s^*)$}

Since the true value $s_T$ can never be determined to infinite
precision, and the true value is the abscissa for which the $pdf$
${\cal P}_n(s^*)$ is the maximum, it follows that the function ${\cal
P}_n(s^*)$ is unknowable. We cannot associate an abscissa to the
function ${\cal P}_n(s^*)$ and hence the function cannot be
``anchored'' to the $s^*$ axis. If we could anchor it, we could 
read off the value of $s_T$ to infinite precision by determining 
the maximum likelihood value of the function. We thus term this function an
``unknown concomitant'', to distinguish it from a Bayesian prior. It
is an abstraction which we approach with ever increasing precision as
we increase $N$, the number of members on the ensemble.
\subsubsection{The standard error on the fitted parameter}
The function ${\cal P}_n(s^*)$ is the probability distribution of the maximum likelihood values $s^*_k$ on the ensemble. The standard error $\sigma_n$ 
of the fitted parameter is defined as
\begin{equation}
\sigma^2_n = \int (s^*-s_T)^2 {\cal P}_n(s^*) ds^* = <(s^*-s_T)^2>
\end{equation}
We also note that ${\cal P}_n(s^*)=<P(s^*|\vec{c_n})>$ is {\it also} the 
ensemble average of the inverse probability functions $P_k(s^*|\vec{c_n})$.
Hence the inverse probability function $P_k(s^*|\vec{c_n})$ from a single 
dataset $k$ is an 
unbiased estimator of the (unknowable) function ${\cal P}_n(s^*)$ and its 
variance can be used to estimate the standard error $\sigma_n$.
\subsubsection{The evaluation of the inverse probability 
$P_k(s^*|\vec{c_n})$: The error bootstrap}
We now need to compute the function $P_k(s^*|\vec{c_n})$. We employ
Bayes' theorem to do this.  The error on the fitted parameter $s^*$
will be related to the width of the inverse probability
$P_k(s^*|\vec{c_n})$ that we are trying to compute. It is also related
to our ignorance of the value of $s_T$ and our inability to anchor the
distribution ${\cal P}_n(s^*)$. Our level of ignorance of where to
anchor the distribution ${\cal P}_n(s^*)$ is directly related to the
error we are trying to compute and is directly related to the width of
$P_k(s^*|\vec{c_n})$.  At this stage, we have worked out the
likelihood ratio ${\cal L_R}^k(s)$ as a function of $s$ and have
evaluated the maximum likelihood value $s^*_k$.  The argument $s$ of
the likelihood ratio is a dummy argument of a function and we are at
liberty to change it to the argument $s^*$ as in ${\cal L_R}^k(s^*)$
for further discussion.  We can choose an arbitrary value of $s^*$ and
evaluate the goodness of fit at that value using the likelihood
ratio. When we do this, we are in fact hypothesizing that $s_T$, the
true value, is at this value of $s^*$. The function ${\cal L}_R(s^*)$
then gives us a way of evaluating the goodness of fit of the
hypothesis as we change $s^*$. Let us now take an arbitrary value of
$s^*$ and hypothesize that that is the true value. Then, consistent
with our hypothesis, we must insist that the distribution ${\cal
P}_n(s^*)$ is moved so that the maximum value of the distribution
(i.e. $s_T$) is at the current value of $s^*$.

\begin{figure}[tbh]
\centerline{\includegraphics[width=\textwidth]{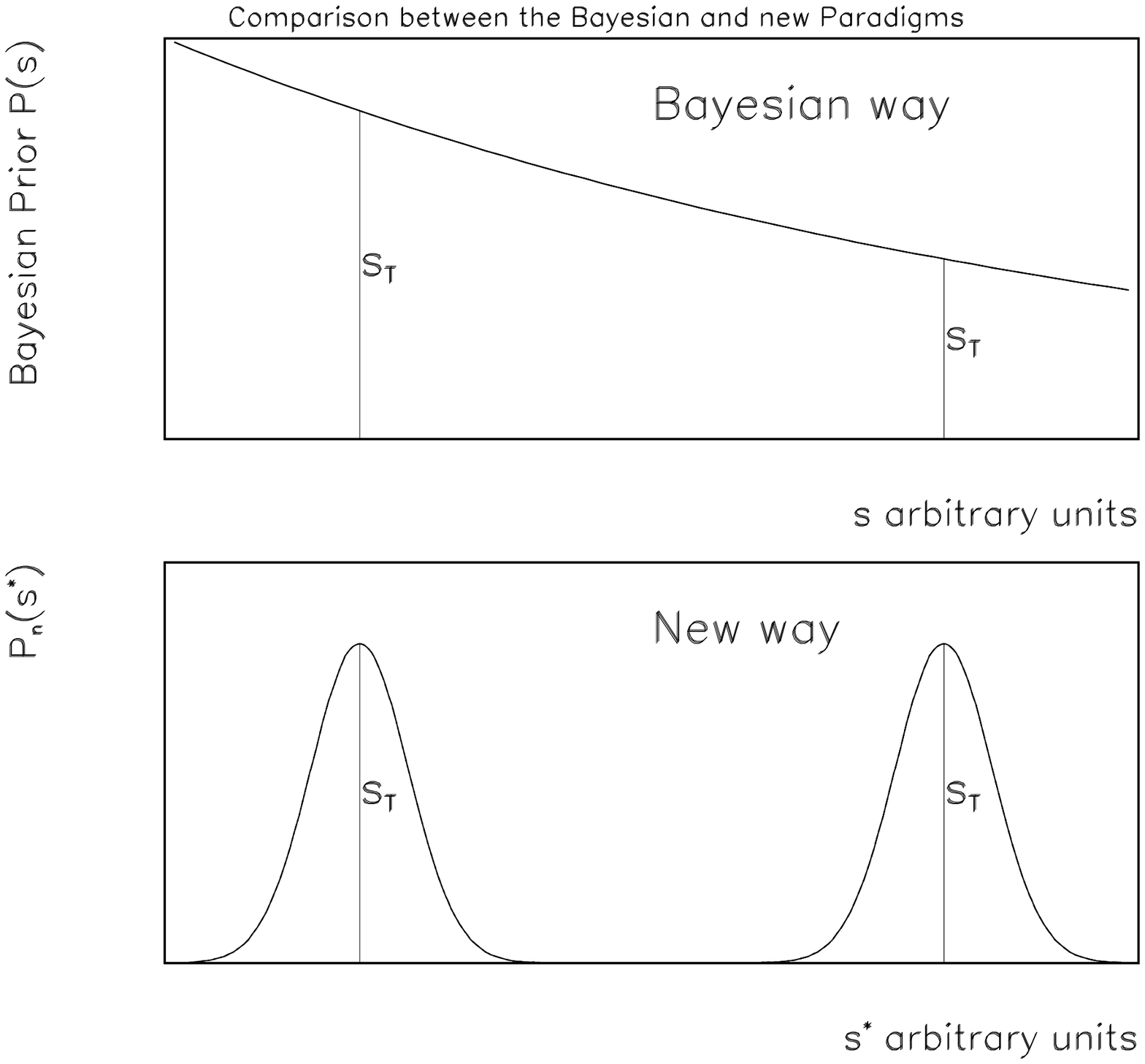}}
\caption[]
{Comparison of the usage of Bayesian priors with the new method. In the upper
figure, illustrating the Bayesian method, 
an unknown distribution is guessed at by the user based on
``degrees of belief'' and the value of the Bayesian prior changes as
the variable $s$ changes. In the lower figure, an ``unknown
concomitant'' distribution ${\cal P}_n(s^*)$ is used whose shape depends on the
statistics of the dataset $\vec{c_n}$. 
In the case of no bias, this distribution peaks at the
true value  $s_T$. As we change $s^*$, we change our hypothesis as to 
where the true value of $s$ lies, and the distribution shifts with $s^*$
as explained in the text. The value of the distribution at the true
value is thus independent of $s^*$.}
\label{bnb}
\end{figure}

At the true value $s_T$, the Bayes' theorem equations for the joint 
probability state
\begin{equation}
 P(s_T,\vec{c_n}) = P(\vec{c_n}|s_T){\cal P}_n(s_T)=P(s_T|\vec{c_n})P^{data}(\vec{c_n}) 
\end{equation}
We now hypothesize that the true value is at $s^*=s_1$. Then the above equation will read
\begin{equation}
 P(s_1,\vec{c_n}) = P(\vec{c_n}|s_1){\cal P}_n(s_T)=P(s_1|\vec{c_n})P^{data}(\vec{c_n}) 
\end{equation}
When we change the hypothesis to a different value $s^*=s_2$, then the equation will read
\begin{equation}
 P(s_2,\vec{c_n}) = P(\vec{c_n}|s_2){\cal P}_n(s_T)=P(s_2|\vec{c_n})P^{data}(\vec{c_n}) 
\end{equation}
We have moved the distribution ${\cal P}_n(s^*)$ to accommodate our
changing hypothesis from the true value being at $s_1$ to $s_2$. These
hypotheses are mutually exclusive in that the true value cannot be at both
$s_1$ and $s_2$. This mandates that we move the function ${\cal P}_n(s^*)$
as we change the hypothesis.
This set of hypotheses thus communicate to our
system of equations, our ignorance of the position of the true
value. The set of hypotheses form an OR of the position of the true
value, whereas by contrast, the Bayesian prior expresses an AND of
the position of the true value.
Then for the hypothesis that the true value is at an arbitrary $s^*$, the above equations become
\begin{equation}
 P(s^*,\vec{c_n}) = P(\vec{c_n}|s^*){\cal P}_n(s_T)=P(s^*|\vec{c_n})P^{data}(\vec{c_n}) 
\label{lreq}
\end{equation}
Re-arranging,
\begin{equation}
  P(s^*|\vec{c_n}) =   
\frac{ P(\vec{c_n}|s^*)}{P^{data}(\vec{c_n})}{\cal P}_n(s_T)
\end{equation}
Imposing the normalization condition $\int P(s^*|\vec{c_n})ds^*=1$ yields
\begin{equation}
  P_k(s^*|\vec{c_n}) =   
\frac{ P_k(\vec{c_n}|s^*)}{\int P_k(\vec{c_n}|s^{*'}) ds^{*'}}
\label{poster}
\end{equation}
where we have explicitly indicated the dependence on the ensemble
index $k$.  To reiterate, when one varies $s^*$ in
equation~\ref{lreq}, one makes the hypothesis that $s=s_T$. As one
changes $s^*$, a new hypothesis is being tested that is mutually
exclusive from the previous one, since the true value can only be at
one location. So as one changes $s^*$, one is compelled to move the
$distribution$ ${\cal P}_n(s^*)$ so that $s_T$ is at the value of
$s^*$ being tested. {\it If one did not move ${\cal P}_n(s^*)$, then this
is tantamount to anchoring the function to the $s^*$ axis and this is
not allowable, since the true value is unknown.} This implies that
${\cal P}_n(s_T)$ does not change as one changes $s^*$ and is a
constant $wrt$ $s^*$. Figure~\ref{bnb} illustrates these points
graphically.  Thus ${\cal P}_n(s_T)$ in our equations is a number, not
a function.
We have thus ``bootstrapped'' the error. On the one hand, 
$P_k(s^*|\vec{c_n})$ gives us an estimate of the spread in the measurements of 
$s^*$ from an ensemble of datasets $\vec{c_n}$, based on one such data set.
On the other hand, the error in $s^*$ is expressed in the
 uncertainty on where to put $s_T$. We have connected these two uncertainties
using Bayes' theorem and hypothesis testing. Also,
\begin{equation}
 {\cal P}_n(s_T) = \frac{1}{\int{\cal L}_R(s^*)ds^*}= 
\frac{P^{data}(\vec{c_n})}{\int P(\vec{c_n}|s^*)ds^*}
\end{equation}
We have thus determined ${\cal P}_n(s_T)$, the value of the ``unknown
concomitant'' at the true value $s_T$ using our data set $c_n$. This
is our $measurement$ of ${\cal P}_n(s_T)$ and different datasets will give
different values of ${\cal P}_n(s_T)$, in other words ${\cal P}_n(s_T)$ will have a
sampling distribution with an expectation value and standard
deviation.

Note that it is only possible to write down an expression
 for ${\cal P}_n(s_T)$ dimensionally when a likelihood ratio ${\cal L_R}$ 
is available. 
The equation~\ref{poster} is the same expression that ``frequentists'' use
for calculating their errors after fitting, namely the likelihood
curve normalized to unity gives the parameter errors. If the
likelihood curve is Gaussian shaped, then this justifies a change of
negative log-likelihood of $\frac{1}{2}$ from the maximum likelihood 
point to get
the $1 \sigma$ errors. Similarly, when performing $\chi^2$ fitting, it
is now rigorously permitted to use $\Delta \chi^2=1$ to estimate
errors in fitted parameters under the Gaussian assumption. No Bayesian
prior is needed.

The ``Objective Bayesians'' may be tempted to remark that
equation~\ref{poster} is the same equation they would derive using a
flat prior and so the two theories are equivalent. This is not the
case, since their projection of the joint probability on the parameter
axis is the Bayesian prior which does not depend on $n$ (for all
Bayesians), whereas in our case it yields a function ${\cal P}_n(s^*)$
which depends on $n$. So the two theories are radically
different. Also, in the new paradigm, one does not have to answer the
question ``Flat in what variable?'', as the ``Objective Bayesians'' have
to do regarding the prior they use. Our theory is invariant no matter
what the density of the hypotheses we make in $s^*$ space is. The
Bayesians will obtain different results when they use different
densities for the prior distribution.
\subsubsection{Iterative behavior of the theory when more than one member of the ensemble is available}

We have now solved the problem for the case when one member of the
ensemble is available. This is what happens in most cases, when only
one dataset exists. If however we want to study the theory when more
the one member of the ensemble is present, we can proceed to use
equation~\ref{lpn} to work out a better approximation for ${\cal
P}_n(s^*)$.
\begin{equation}
 {\cal P}_n(s^*) = <P(s^*|\vec{c_n})> \approx \frac{1}{N}\sum_{k=1}^{k=N} P_k(s^*|\vec{c_n})
\label{lpn1}
\end{equation}
We now have an approximation to the function ${\cal P}_n(s^*)$ which
is based on $N$ datasets instead of one. This approximation can be
used to iterate our functions $P_k(s^*|\vec{c_n})$ using
equation~\ref{invnew}.
\begin{equation}
 P^{(2)}_k(s^*|\vec{c_n}) = 
\frac{P_k(\vec{c_n}|s^*){\cal P}_n(s^*)}
{\int P_k(\vec{c_n}|s^{*'}){\cal P}_n(s^{*'})ds^{*'}}
\label{iter1}
\end{equation}
where we have used the superscript $(2)$ to indicate that this is the second iteration of the function. These sets of functions can be used to obtain a new version of the function ${\cal P}_n(s^*)$ using the relation
\begin{equation}
 {\cal P}^{(2)}_n(s^*) =  \frac{1}{N}\sum_{k=1}^{k=N} P_k^{(2)}(s^*|\vec{c_n})
\label{iter2}
\end{equation}
Similarly, one needs to iterate on the theoretical likelihoods 
$P_k(\vec{c_n})|s^*)$ when information from more than one member of 
the ensemble is available. This is done by using equation~\ref{lpc} 
to obtain an approximation for ${\cal P}(\vec{c_n})$.
\begin{equation}
{\cal P}(\vec{c_n})=<P(\vec{c_n}|s^*)>
\approx \frac{1}{N}\sum_{k=1}^{k=N}P_k(\vec{c_n}|s^*)\\
\end{equation}
This equation states that it is possible to get a better estimate of
the density of data ${\cal P}(\vec{c_n})$ from all the $N$ members of
the ensemble combined than from a single member alone. This is followed by
\begin{equation}
 P^{(2)}_k(\vec{c_n}|s^*) = 
\frac{P_k(s^*|\vec{c_n}){\cal P}(\vec{c_n})}{\int P_k(s^*|\vec{c_n^{'}})
{\cal P}(\vec{c_n^{'}})d\vec{c_n^{'}}}
\label{iter3}
\end{equation}
This equation states that in the presence of all $N$ members of the
ensemble, it is possible to obtain a better value of the likelihood
$P_k(\vec{c_n}|s^*)$ than the theoretical likelihood which assumes
that the true value $s_T$ is at $s^*$.
The above functions are used to derive an iterated version of the data density
${\cal P}(\vec{c_n})$ that uses information available from all the members of 
the ensemble to compute the data density.
\begin{equation}
 {\cal P}^{(2)}(\vec{c_n}) =  \frac{1}{N}\sum_{k=1}^{k=N} P_k^{(2)}(\vec{c_n}|s^*)
\label{iter4}
\end{equation}

\subsubsection{Combining Results of Experiments}
Each experiment should publish a likelihood curve for its fit as well as a 
number for the data likelihood $P^{data}(\vec{c_n})$. Combining the results of 
two experiments with $m$ and $n$ experiments each, involves multiplying the 
likelihood ratios.
\begin{equation}
  {\cal L_R}\:_{m+n}(s) ={\cal L_R}\:_m(s) \times {\cal L_R}\:_n(s) = 
\frac{P(\vec{ c_m}|s)}{P^{data}(\vec{ c_m})} 
\times \frac{P(\vec{c_n}|s)}{P^{data}(\vec{c_n})}
\end{equation}
Inverted probabilities and goodness of fit can be deduced from the
combined likelihood ratio. It is worth noting that the numerator of
the likelihood ratio is a function of the fitted parameters and the
denominator is a number.
\subsubsection{Another Illustrative Example}
\label{aillex}
We now apply the theory developed here to a practical example to 
illustrate the ideas further. 
The problem is to determine the weight of an object using an apparatus
whose standard error is known to be 5~gm. The weight is a fixed constant 
of nature for the duration of the experiment. We obtain a  dataset
of 100 measurements, i.e. $n=100$. Then $P(c|s)$ is a Gaussian of
unknown mean s and  width $\sigma=5$~gm. We compute
$P(\vec{c_n}|s)$ for the 100 events by multiplying the individual
$P(c_i|s)$ together and maximize the likelihood to determine $s^*$ for
the dataset using unbinned likelihoods. We then transform the measurements 
$c_i$ to the hypercube space using equation~\ref{hyp}. We use the improved 
$PDE$ in hypercube space with $h=0.2$ and determine the goodness of fit and 
the negative log-likelihood ratio ${\cal NLLR}$. We repeat this for an ensemble
of 1000 experiments.
\begin{figure}[tbh]
\centerline{\includegraphics[width=\textwidth]{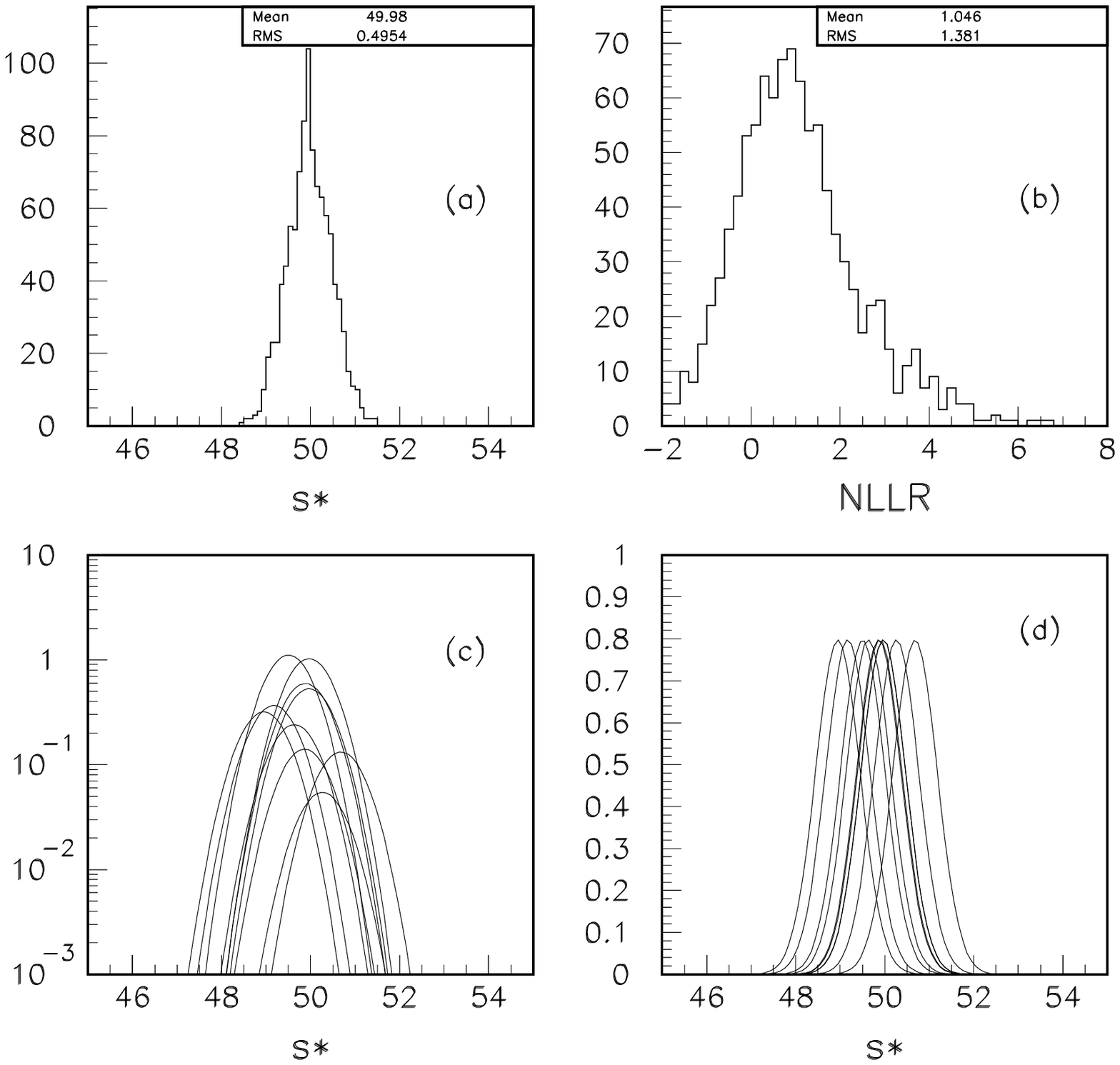}}
\caption[]
{(a)The distribution of $s^*$, the maximum likelihood value of $s$ for
a 1000 member ensemble of datasets of $n=100$. (b)The goodness of fit variable
${\cal NLLR}$ for the fits (c)The likelihood ratio ${\cal L}_R(s^*)$ as a function of $s^*$ for the first 10  members of the ensemble (d)
The function $P(s^*|\vec{c_n})$ for the first 10 members of the ensemble.}
\label{ensem}
\end{figure} 

Figure~\ref{ensem}(a) shows the distribution of $s^*$ for this
ensemble. The mean value of $s^*$ over this ensemble is 49.98~gm and
the RMS is 0.495~gm which is consistent with the expected
$\sigma/\sqrt(100)$ value of 0.5~gm. Figure~\ref{ensem}(b) shows the
distribution of ${\cal NLLR}$ for the 1000 members of the
ensemble. Figure~\ref{ensem}(c) shows the likelihood ratio functions
${\cal L}_R(s^*)$ for the first 10 fits in the ensemble. The value of
$s^*_k$, the maximum likelihood value of the $k^{th}$ member of the ensemble 
fluctuates as expected, as well as the value of ${\cal
L}_R(s^*_k)$, the negative logarithm of which gives the ${\cal
NLLR}$. The fluctuation in $s^*_k$ for the fits in the ensemble
essentially expresses our lack of knowledge of the position of the
true value $s_T$. The width of the likelihood distribution also
contains information on the same lack of knowledge.

We now use equation~\ref{poster} to obtain inverse probabilities $P_k(s^*|\vec{c_n})$ for each member of the ensemble.  These functions are shown in
Figure~\ref{ensem}(d). The maximum likelihood value moves around with the 
expected spread of 0.5~gm. The average standard deviation of  these curves is 
0.5~gm with an rms of  0.65~E-3~gm.
The average of these functions on an infinite ensemble yields the true
$pdf$ ${\cal P}_n(s^*)$.

\subsubsection{Iterative behavior of the theory in the example} 
In practice, if one has a dataset with $n=100$ and $N=1000$ similar
instances of them, the easiest way to analyze the data is to combine them all
into a dataset with $n'=Nn=100,000$. However, we are interested in studying
the function ${\cal P}_n(s^*)$ which is estimated by the ensemble
average of the functions $P_k(s^*|\vec{c_n})$. This function tells us the 
behavior of the distribution of the maximum likelihood values $s^*$ over 
similar datasets each with n=100.

\begin{figure}[tbh]
\centerline{\includegraphics[width=\textwidth]{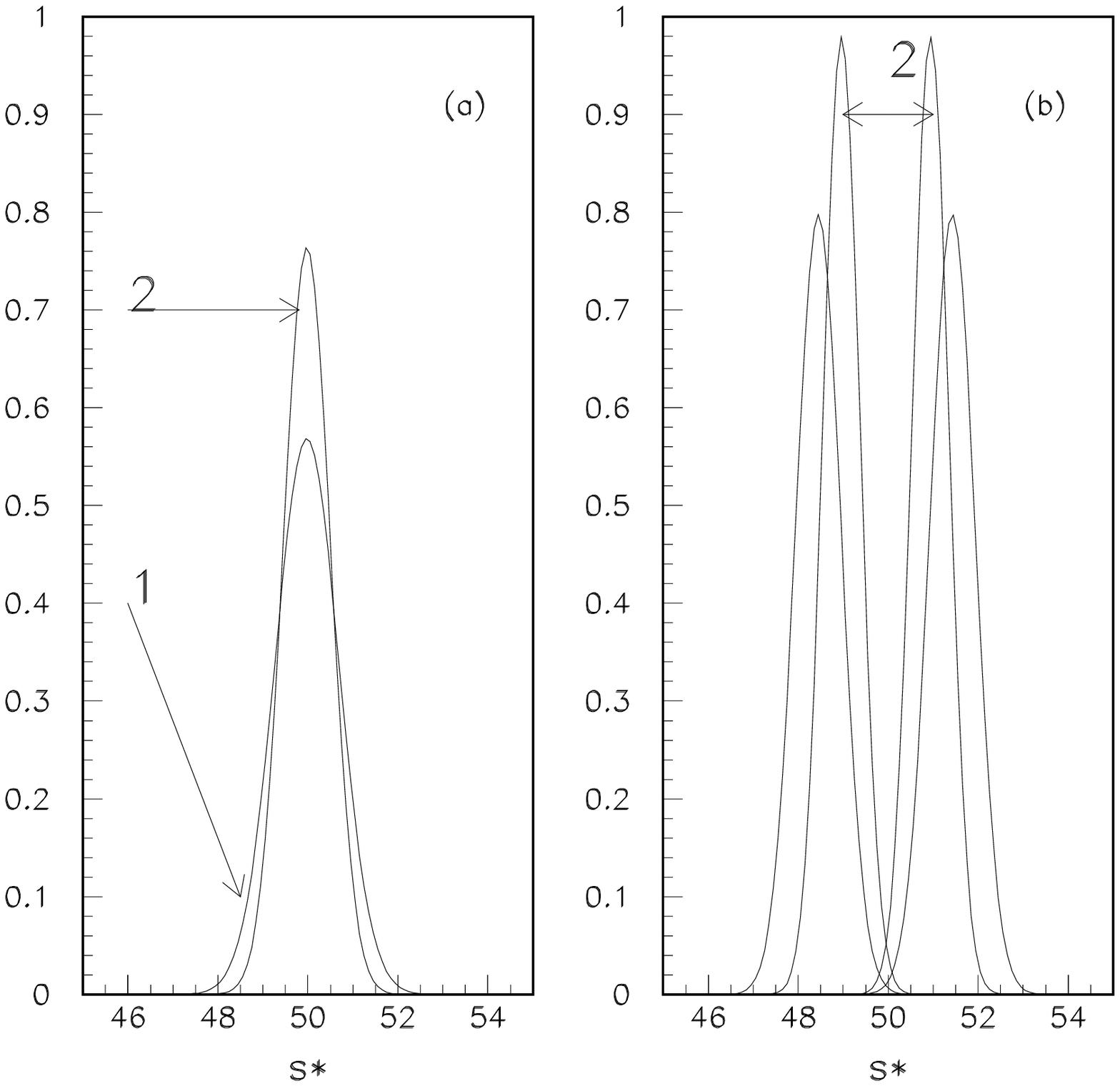}}
\caption[]
{(a) The function ${\cal P}_n(s^*)$ computed on the ensemble for n=100
and N=1000. The two iterations are shown, with the numbers (1,2) indicating 
the iteration number. (b) The function
$P(s^*|\vec{c_n})$ for two elements on the ensemble for the two
iterations.}
\label{ensiter}
\end{figure}

\begin{figure}[tbh]
\centerline{\includegraphics[width=\textwidth]{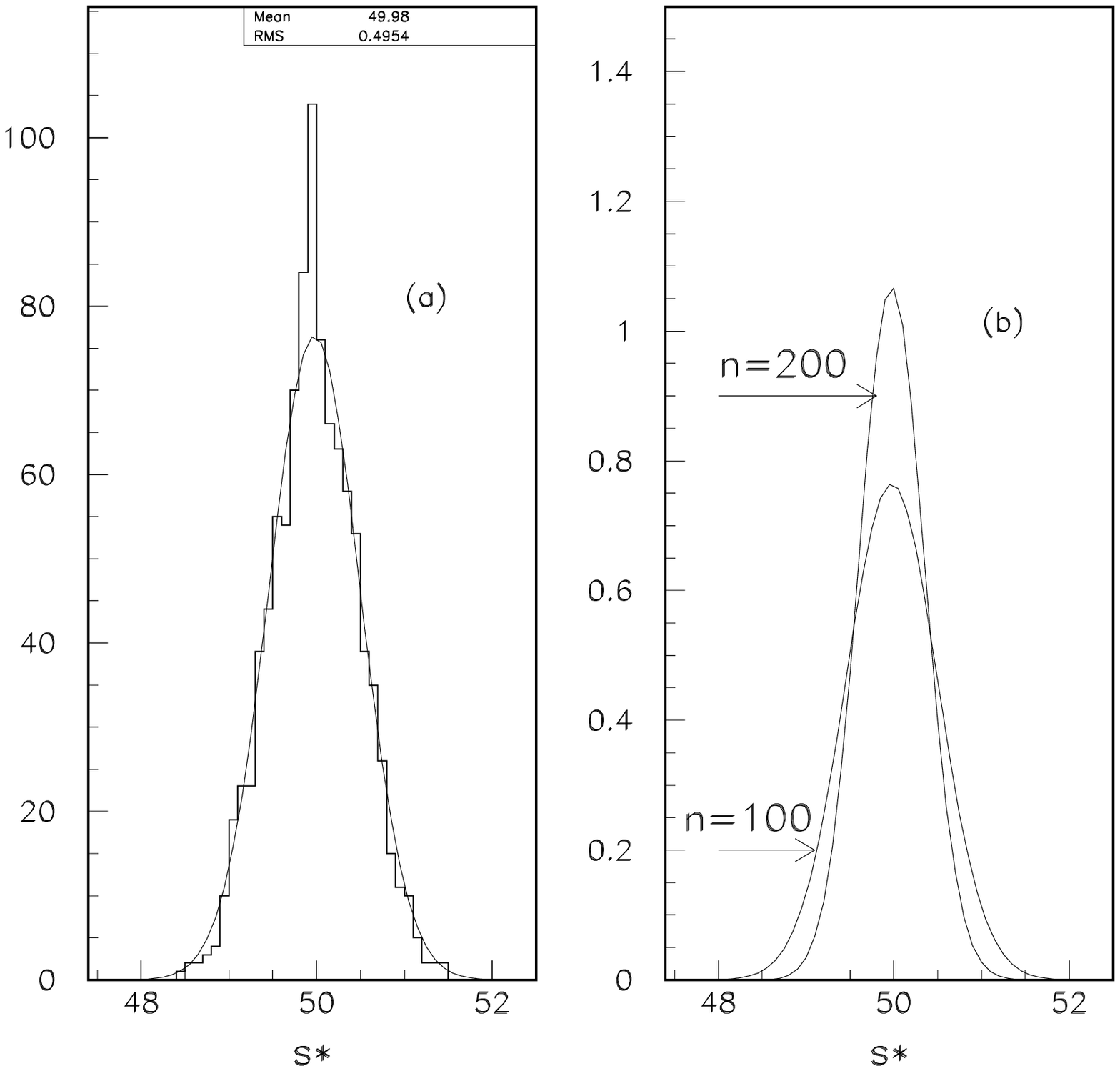}}
\caption[]
{(a)The distribution of $s^*$ (solid histogram) for an ensemble with N=1000 
elements each consisting of a dataset n=100. The curve is the estimate for the
iterated function ${\cal P}_n(s^*)$ for this ensemble normalized to 
the 1000 observations. (b) ${\cal P}_n(s^*)$ on the ensemble for n=100 and
 n=200. This illustrates that the ensemble averaged function, 
depends on $n$, the size of the dataset. As $n$ increases, the function 
narrows and the value of the function at its maximum increases. }
\label{psnorm}
\end{figure}

We now iterate to re-determine 
$P_k(s^*|\vec{c_n})$ and ${\cal P}_n(s^*)$ as 
per equations~\ref{iter1} and~\ref{iter2}. 
Figure~\ref{ensiter}(a) shows the ensemble average 
estimate of ${\cal P}_n(s^*)$ for n=100 and N=1000 before and after iteration.
The mean value of the un-iterated and iterated functions are the same at 
49.977~gm (The Gaussians were generated with a true value of 50~gm).
The r.m.s values  of the function  before and after iteration are
 0.701~gm and 0.522~gm respectively. 
The iterated function thus has the correct width and mean value. 
Figure~\ref{ensiter}(b) shows the individual $P_k(s^*|\vec{c_n})$ 
functions for two members of the ensemble before and after iteration. 
The iterations pull these functions towards the true value, since we are 
inputing additional information on the true value.
Figure~\ref{psnorm}(a) shows the values of $s^*$ histogrammed for our 
illustrative example for an ensemble of N=1000 and n=100. 
The superimposed curve is the iterated  function ${\cal P}_n(s^*)$ calculated 
for this ensemble normalized to a 1000 element ensemble. It can be seen that 
the  function describes the distribution of $s^*$ well. 
Figure~\ref{psnorm}(b) shows the iterated function ${\cal P}_n(s^*)$ 
for $n=100$ and $n=200$ respectively. As expected, the $n=200$ 
function is narrower and its value at the maximum is larger, 
illustrating that ${\cal P}_n(s_T)$ increases with $n$.

\begin{figure}[tbh!]
\includegraphics[width=\textwidth]{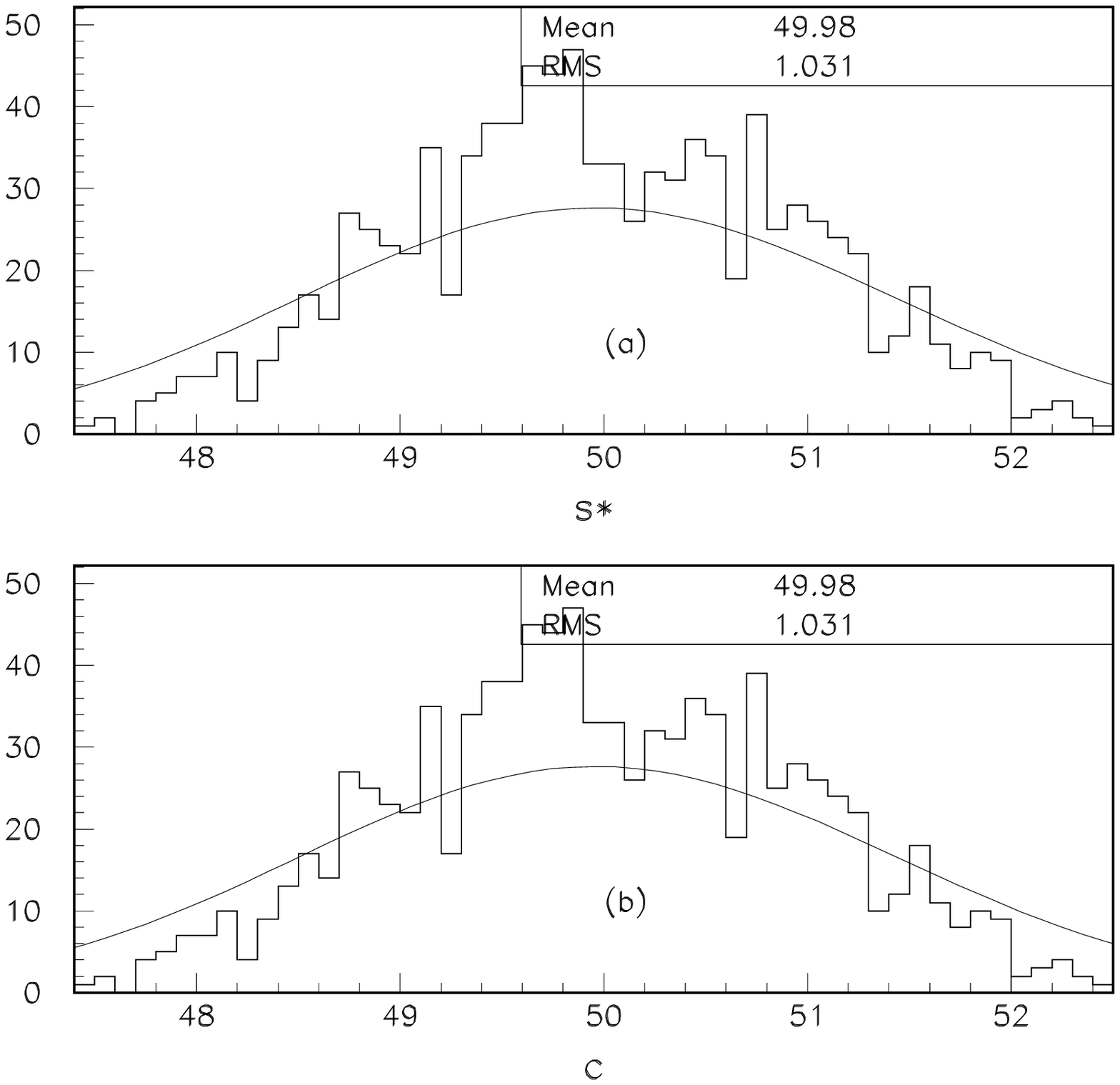}
\caption
{(a)Histogram of $s^*$ values over an ensemble of N=1000. Superimposed
is our first iteration of the function ${\cal P}_n(s^*)$. (b)
Histogram of $c$ values over an ensemble of N=1000. Superimposed is our 
first iteration for the function $P(c)$. The RMS values refer to the width of the histogram. The first iteration curves are too wide as explained in the text.
\label{first}}
\end{figure}

\begin{figure}[tbh!]
\includegraphics[width=\textwidth]{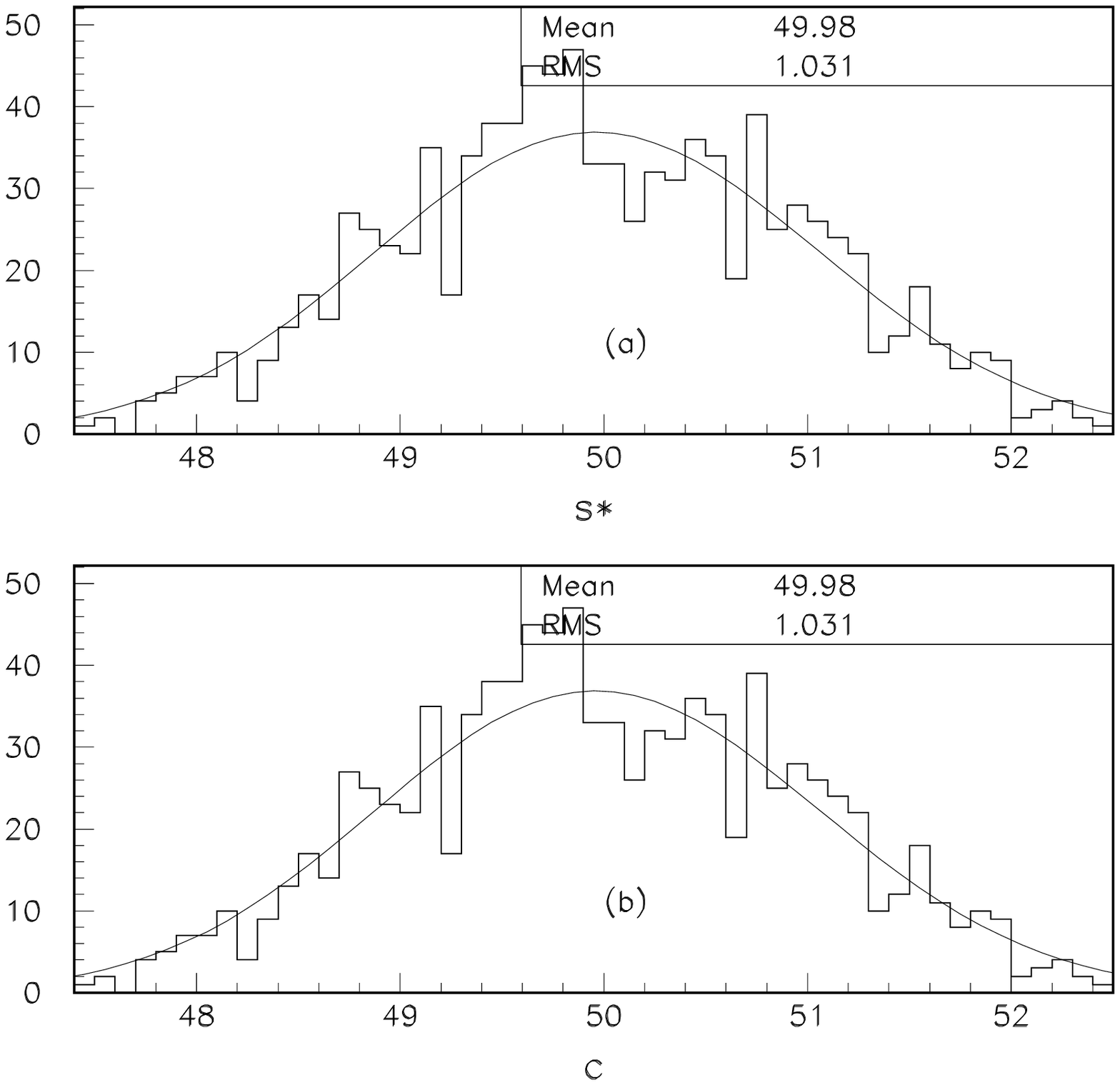}
\caption
{(a)Histogram of $s^*$ values over an ensemble of N=1000. Superimposed
is our second iteration of the function ${\cal P}_n(s^*)$. (b)
Histogram of $c$ values over an ensemble of N=1000. Superimposed is our 
second iteration for the function $P(c)$.
\label{second}}
\end{figure}

\begin{figure}[tbh!]
\includegraphics[width=\textwidth]{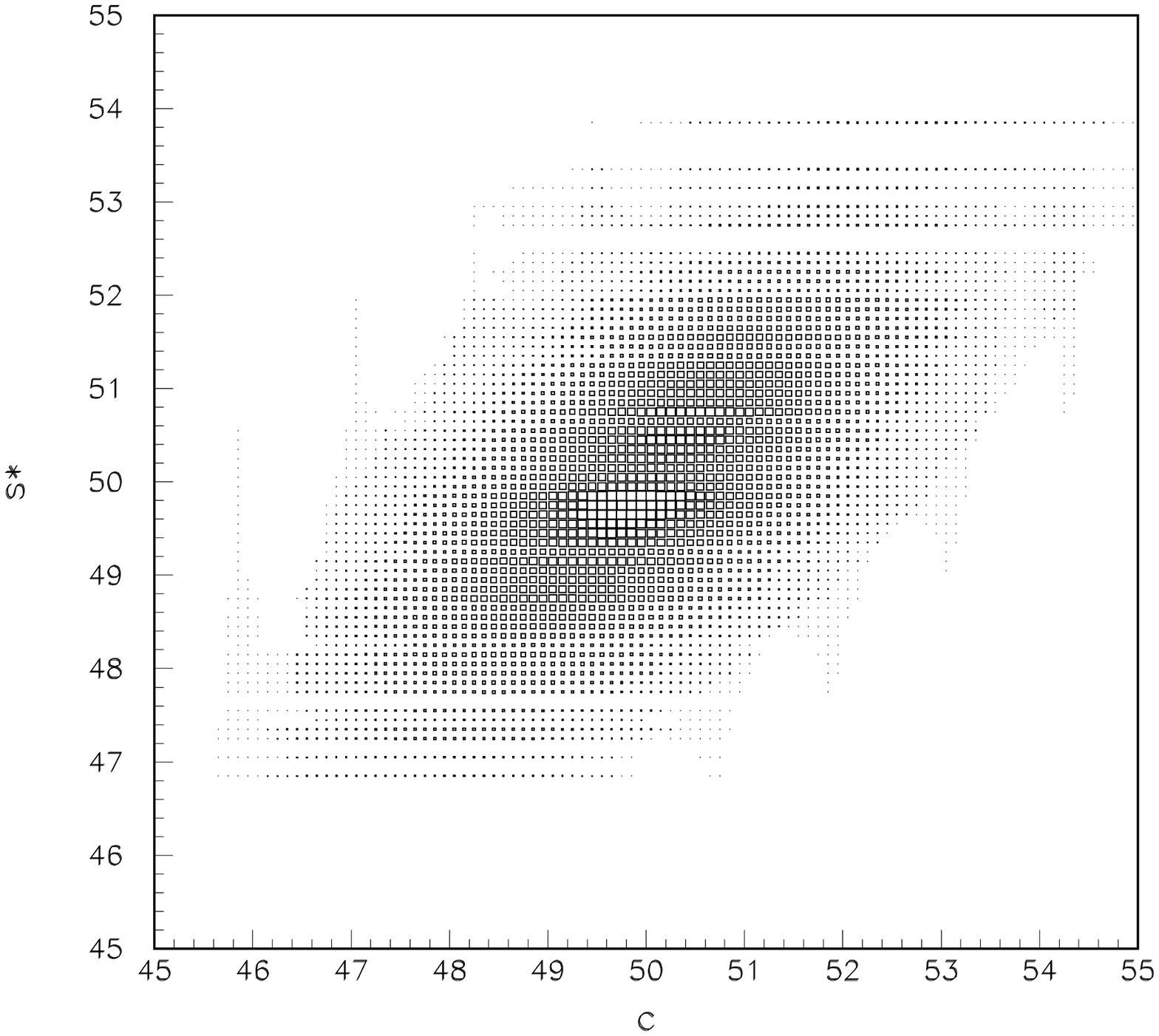}
\caption
{Joint probability $P(c,s^*)$ computed from $P(c|s^*)P(s^*)$ at the end of 
two iterations
\label{jointa}}
\end{figure}

\begin{figure}[tbh!]
\includegraphics[width=\textwidth]{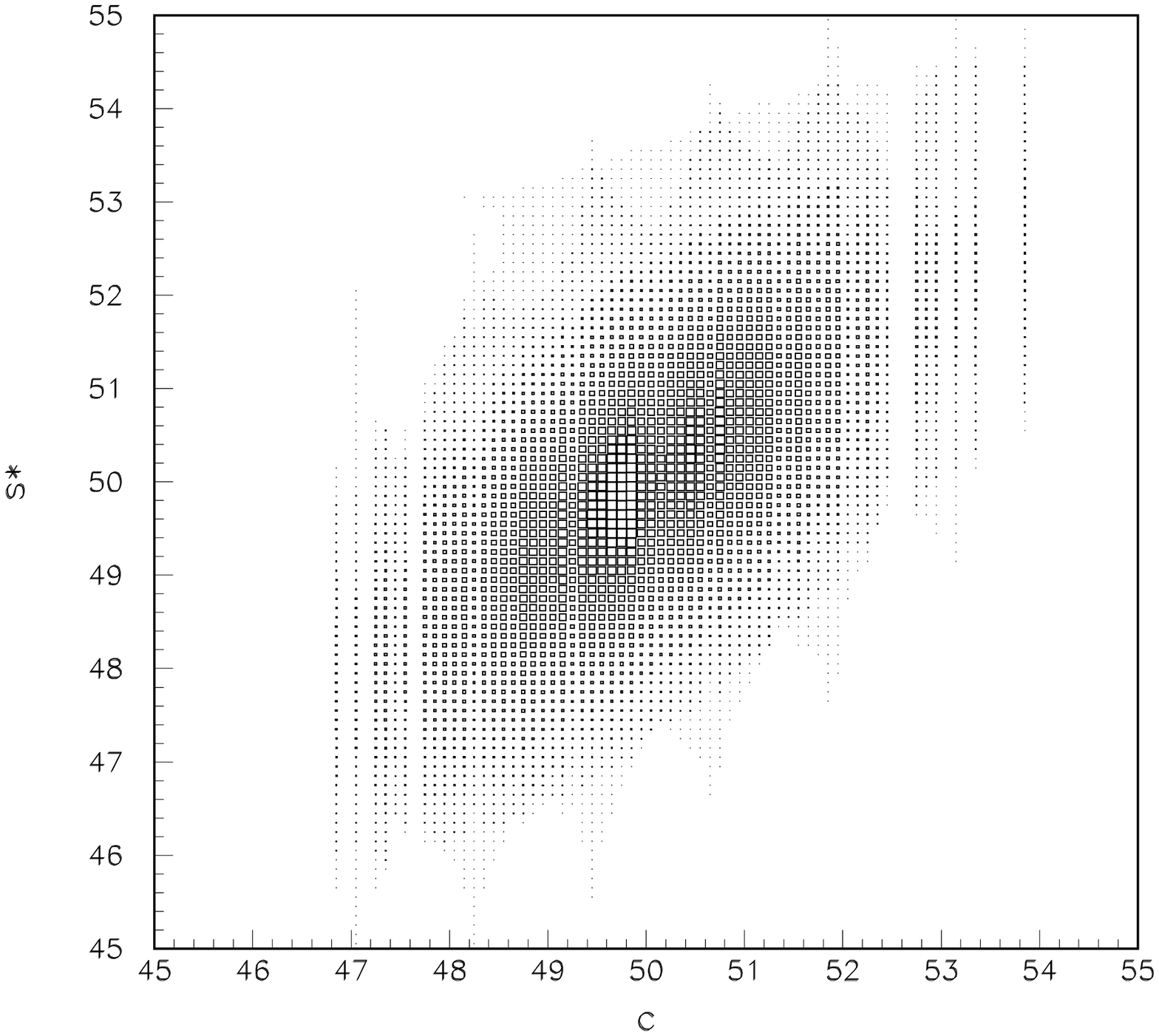}
\caption
{Joint probability $P(c,s^*)$ computed from $P(s^*|c)P(c)$ at the end of 
two iterations
\label{jointb}}
\end{figure}

We can proceed to iteratively work out the likelihoods 
$P_k(\vec{c_n}|s^*)$ and $P(\vec{c_n})$ as per 
equations~\ref{iter3} and ~\ref{iter4}. However, it is difficult to plot
these functions since their argument is multidimensional. 
Instead we show how the iteration works for a special case of the 
above example 
where the dataset consists of a single measurement, i.e. $n=1$. 
We consider an ensemble of $N=1000$ measurements each with a $\sigma =1.0$~gm.
Each single measurement $c$ is fitted to a Gaussian likelihood. The maximum 
likelihood point $s^*$ is trivially equal to $c$ and the goodness of fit likelihood ratio is always unity. The inverted probability is a Gaussian given by
\begin{equation}
 P(s^*|c) = \frac{e^{-(s^*-c)^2/2\sigma^2}}{\sqrt{2\pi}\sigma}
\end{equation}
We proceed to work out the function ${\cal P}_n(s^*)$ by averaging the
above functions over the ensemble. The resulting Gaussian will be a
convolution of two Gaussians with width $\sigma=1.0$ and will possess
a width $\sigma$ = $\sqrt{2}$. Similarly we proceed to work out ${\cal
P}(c)$ by averaging the theoretical likelihoods $P_k(c|s^*)$ over the
ensemble. This curve will also be too wide for similar
reasons. Figure~\ref{first} shows the resulting curves for the first
iteration plotted on top of the histograms for the data for $s^*$ and
$c$ respectively. Figure~\ref{second} shows the curves after the
second iteration and both the curves fit well.
We now plot the resulting joint probability $P(s^*,c)$ obtained two
different ways. Figure~\ref{jointa} shows the joint probability worked
out by the Bayes' theorem equation $P(s^*,c)= P(c|s^*){\cal P}_1(s^*)$ and
Figure~\ref{jointb} shows the joint probability worked out by the
equation $P(s^*,c) = P(s^*|c){\cal P}(c)$ after the iterations have been
made. It can be seen that both these procedures give the same joint
probability distribution that possesses a correlation between $c$ and
$s^*$ that is less extreme than the initial correlation of
$c=s^*$. The projections of the joint probability on the $c$ and the
$s^*$ axes fit the data well. We have iteratively solved Bayes'
theorem on the ensemble and inverted the probability correctly without
the use of a Bayesian prior.
\subsection{The two different methods to obtain ${\cal P}_n(s^*)$}
In our theory, ${\cal P}_n(s^*)$ is the function obtained by
histogramming the maximum likelihood values $s^*_k$ for an infinite
ensemble of datasets $\vec{c_n}$ and normalizing the resulting
histogram to unity. i.e. it is the probability density function of the
maximum likelihood values on the ensemble, for datasets each containing 
$n$ elements. However, equation~\ref{lpn1} shows another way of 
obtaining the same function. What is the connection between the two methods?

Without loss of generality, we can 
express the inverse probability  function as a function of $s^*-s^*_k$ 
such that
\begin{equation}
 P_k(s^*|\vec{c_n}) \equiv {\cal G}_k(s^*-s^*_k)
\end{equation}
Then equation~\ref{lpn1} can be re-expressed
\begin{equation}
{\cal P}_n(s^*)= 
\lim_{N\rightarrow \infty} \frac{1}{N}\sum_{k=1}^{k=N} {\cal G}_k(s^*-s^*_k)
\end{equation}
But this is just the probability density estimator ($PDE$)  
for the distribution of $s^*$, with 
the functions ${\cal G}_k$ serving as the kernels!. They satisfy the 
normalization condition $\int {\cal G}_k(t) dt =1$ as required. This should 
be compared with equation~\ref{pde} for the definition of $PDE's$. 
Thus ${\cal P}_n(s^*)$ represents a $PDE$ of the distribution of $s^*$ and 
will yield the same distribution as $s^*$.

In the limit $N\rightarrow \infty$, we can represent the distribution of 
the maximum likelihood values $s^*$ on the ensemble as 
the continuous $pdf$ ${\cal P}_n(s^*)$.
In this limit,  one can write
\begin{equation}
 {\cal P}_n(s^*) = \int {\cal P}_n(s^{*'}) {\cal G}(s^{*'}, s^*-s^{*'}) ds^{*'}
\label{inteq}
\end{equation}

where we have used the notation ${\cal G}(s^{*'},s^*-s^{*'})$ to
emphasize the variation of the kernel as a function of $s^{*'}$
(i.e. ensemble element).  The latter half of the above equation is an
integral equation with kernel ${\cal G}(s^{*'},s^*-s^{*'})$ whose
eigenfunction is ${\cal P}_n(s^*)$.  

Let us also note that the iterative method used to solve Bayes'
theorem in the example given above where $c=s^*$, can be used as a PDE
method to adjust the kernels by changing their shape iteratively
without resort to an adjustable parameter $h$. We could have fed in
data generated as an exponential(for example) with the assumption that
each measurement has a Gaussian error. Then each of the Gaussian
kernels would have been altered by the resulting exponential 
function ${\cal P}_n(s^*)$
iteratively yielding a PDE for the exponential.

\subsection{Co-ordinate transformations $s^{*'}=s^{*'}(s^*)$}

The inverse probability density functions $P(s^*|\vec{c_n})$ are 
invariant under the co-ordinate transformations $c'=c'(c)$. 
How do they behave under  transformations  $s{*'}=s^{*'}(s^{*})$?  The function
$P_k(s^*|\vec{c_n})$ represents our estimate using one member of 
the ensemble of the $pdf$ of $s^*$. So if $P_k(s^*|\vec{c_n})$ 
represents a $pdf$, 
we would expect it to behave like a $pdf$, namely
\begin{equation}
 P_k(s^{*'}|\vec{c_n}) = P(s^*|\vec{c_n}) |\frac{\partial s^*}{\partial s^{*'}}|
\end{equation}
This is how $pdf's$ transform (via the Jacobian). This can be shown patently 
not to be so, since
$P_k(\vec{c_n}|s^{*'})=P_k(\vec{c_n}|s^*)$ and
\begin{equation}
 P_k(s^{*'}|\vec{c_n}) = 
\frac{P_k(\vec{c_n}|s^{*'})}{\int P_k(\vec{c_n}|s^{*'})ds^{*'}} 
 = \lambda_k(\vec{c_n})  P_k(s^*|\vec{c_n})
\end{equation}
where the $s^*$ independent constant $\ \lambda_k(\vec{c_n})$ is given by
\begin{equation}
 \lambda_k(\vec{c_n}) = \frac{ \int P_k(\vec{c_n}|s^*)ds^*}
{\int P_k(\vec{c_n}|s^{*'})ds^{*'}}
\end{equation}
i.e. the inverse probability 
densities do not transform in a way that is expected of
 $pdf's$. This was perhaps a naive expectation. 
As we have just demonstrated, the inverse probability densities serve the
purpose of kernels on the ensemble, the ensemble average of which gives
the $pdf$ ${\cal P}_n(s^*)$. There is no need for the kernel from a
member of the ensemble to transform to the kernel from the same member
under these transformations. The properties of the ensemble average
deduced from the individual kernels will fluctuate from kernel to
kernel. Similarly, when one analyzes in transformed variables, the
same kernel will give different results which may be thought of as
being part of the  fluctuation. 

The distributions of the maximum 
likelihoods ${\cal P}_n(s^*)$ however will transform as $pdf's$, 
since ${\cal P}_n(s^*)$ 
represents the probability density of the maximum likelihood value and 
$s'^*= s'(s^*)$. i.e.
\begin{equation}
 {\cal P}_n^{'}(s^{*'}) = |\frac{\partial s^*}{\partial s^{*'}}| {\cal P}_n(s^*)
\label{ggg}
\end{equation}
The transformed kernels {\it after iteration} will yield the transformed  
${\cal P}_n^{'}(s^{*'})$.
\subsubsection{Comparison with the Bayesian approach}
Table~\ref{diff} outlines the major differences between the
Bayesian approach and the new paradigm.
\begin{table}[bht!]
\caption[Differences between the two methods]{
The key points of difference between the Bayesian method and the new method.
\label{diff}}
\centering\leavevmode
\begin{tabular}{|l|l|l|}
\hline
Item & Bayesian Method & New Method \\
\hline
Goodness & Absent & Now available  \\
 of fit  &        & in both binned \\
         &        & and unbinned fits\\
\hline
Data & Used in evaluating  & Used in evaluating \\
     &       theory $pdf$  &  theory $pdf$\\
     &  at data points     &  at data points \\
     &                     & as well as evaluating \\
     &                     &data $pdf$ at data points\\
\hline
Prior & Is a distribution   & No prior needed. \\
      &that is guessed based & One calculates a \\
      &on ``degrees of belief'' & constant from data\\
      &Independent of data, & 
 ${\cal P}_n(s_T) = \frac{P^{data}(\vec{c_n})}{\int P(\vec{c_n}|s^*) ds^*}$ \\
      &monolithic           & $\rightarrow \infty$ as $n\rightarrow \infty$ \\
\hline 
Inverse probability & Depends on Prior. &  Independent of prior. \\
density & & same as frequentists use  \\
 & 
$P(s|\vec{c_n})=\frac{P(\vec{c_n}|s)P(s)}{\int P(\vec{c_n}|s')P(s')~ds'}$ 
& $P(s^*|\vec{c_n})=\frac{P(\vec{c_n}|s^*)}{\int P(\vec{c_n}|s')~ds'}$  \\
\hline
\end{tabular}
\end{table}
\section{Conclusions}

To conclude, we have proposed a general theory for obtaining the
goodness of fit in likelihood fits for both binned and 
unbinned likelihood fits. In order to obtain a goodness of fit measure,
one needs two likelihoods:- one derived from theory and the other
derived from the data alone. In order to compute the errors on fitted
quantities, inverse probability densities need to be worked out and Bayes'
theorem needs to be employed. Using insights gained in 
solving the goodness of fit problem, we demonstrate that 
it is possible to estimate the inverse probability densities 
without the use of Bayesian prior.
This results in a new paradigm in statistics.

\section{Acknowledgments}

This work is supported by Department of Energy.

\section{Appendix}

In order to demonstrate the capabilities of the unbinned goodness of 
fit method, we illustrate its power with the following example.

\subsection {An extreme problem}
We now attempt to solve a problem with three observed data points, made extreme due to the sparsity of data. The problem is stated as follows.

``Three data points are observed~\cite{knuteson} in three dimensional co-ordinate 
space x,y,z with 
 (x,y,z) = (0.1,0.2,0.3), (0.2,0.4,0.1), and (0.05,0.6,0.21).
What is the goodness of fit to the hypothesis that  the observed
number of events is distributed according to $p(x,y,z)=e^{-(x+y+z)}$ ?''

\subsection{Goodness of fit for the above problem}

We note that the likelihood function for the problem is
\begin{equation}
{\cal L} = \prod_{i=1}^{i=3} \frac{1}{s}exp-\left((x_i+y_i+z_i)/s\right)
\end{equation}
where we assume a maximum likelihood fit has been done and
the lifetime parameter $s$ has been determined to be $s^*=1$ at the maximum.
Since the three co-ordinates x,y, and z are uncorrelated (as per the above likelihood function), we can reformulate the problem as a single dimensional problem as follows.

\begin{equation}
{\cal L} = \prod_{i=1}^{i=9} \frac{1}{s}exp\left(-c_i/s\right)
\end{equation}
where the n=9 vector $\vec{c_n} =0.1\:0.2\:\:0.3\:\:\:0.2\:\:0.4\:\:0.1\:\:\:0.05\:\:0.6\:\:0.21$

We transform the co-ordinates to the hypercube space ($s^*=1$), with the 
limits of $c$ assumed to be 0.0 and 10.0
\footnote{Since the program expects a finite upper limit, 
the high value of c=10 is deemed to be sufficiently large to 
be infinite for this problem. }.

\begin{figure}[tbh!]
\includegraphics[width=\textwidth]{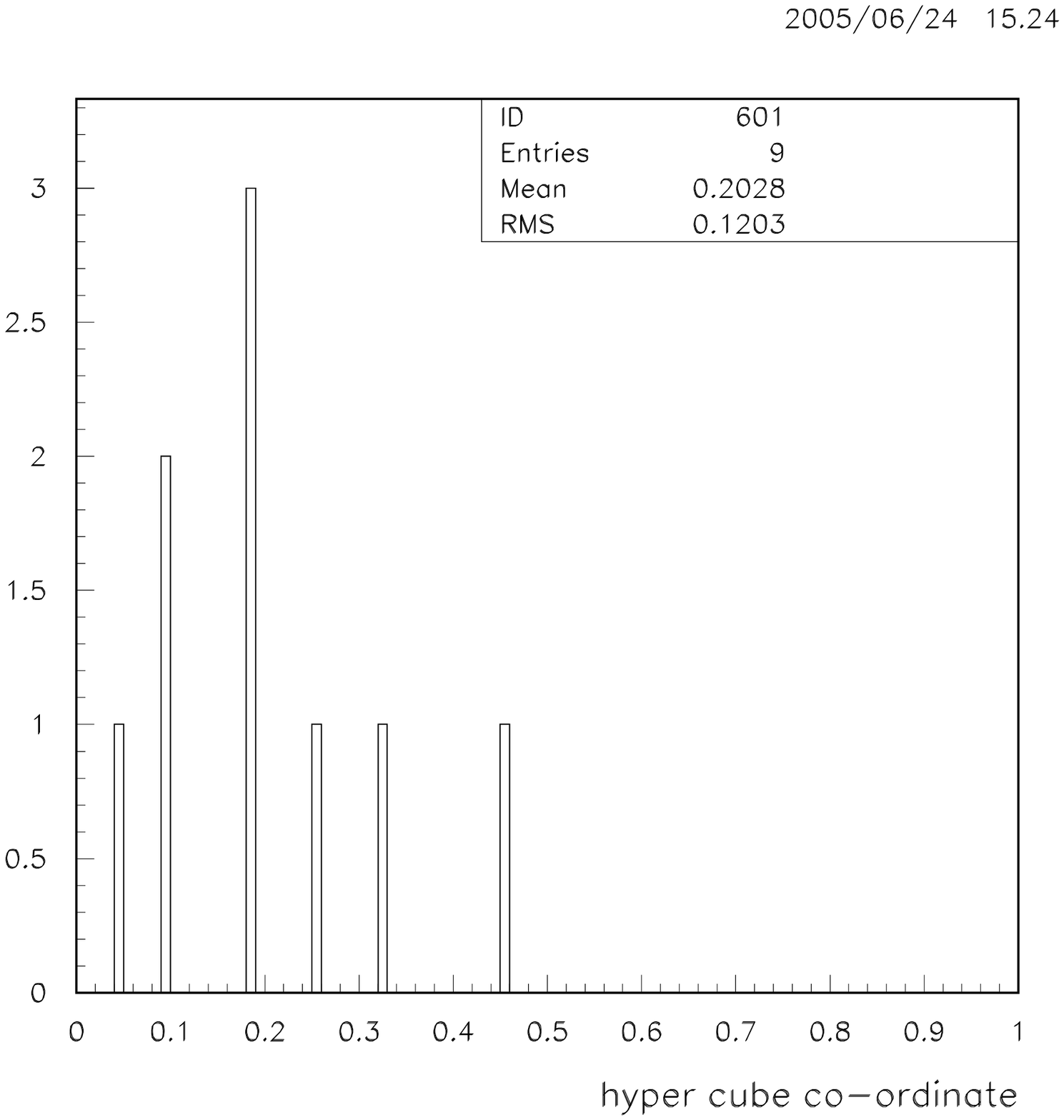}
\caption
{Transformed co-ordinates in hypercube space.
\label{hyper}}
\end{figure}

\begin{figure}[tbh!]
\includegraphics[width=\textwidth]{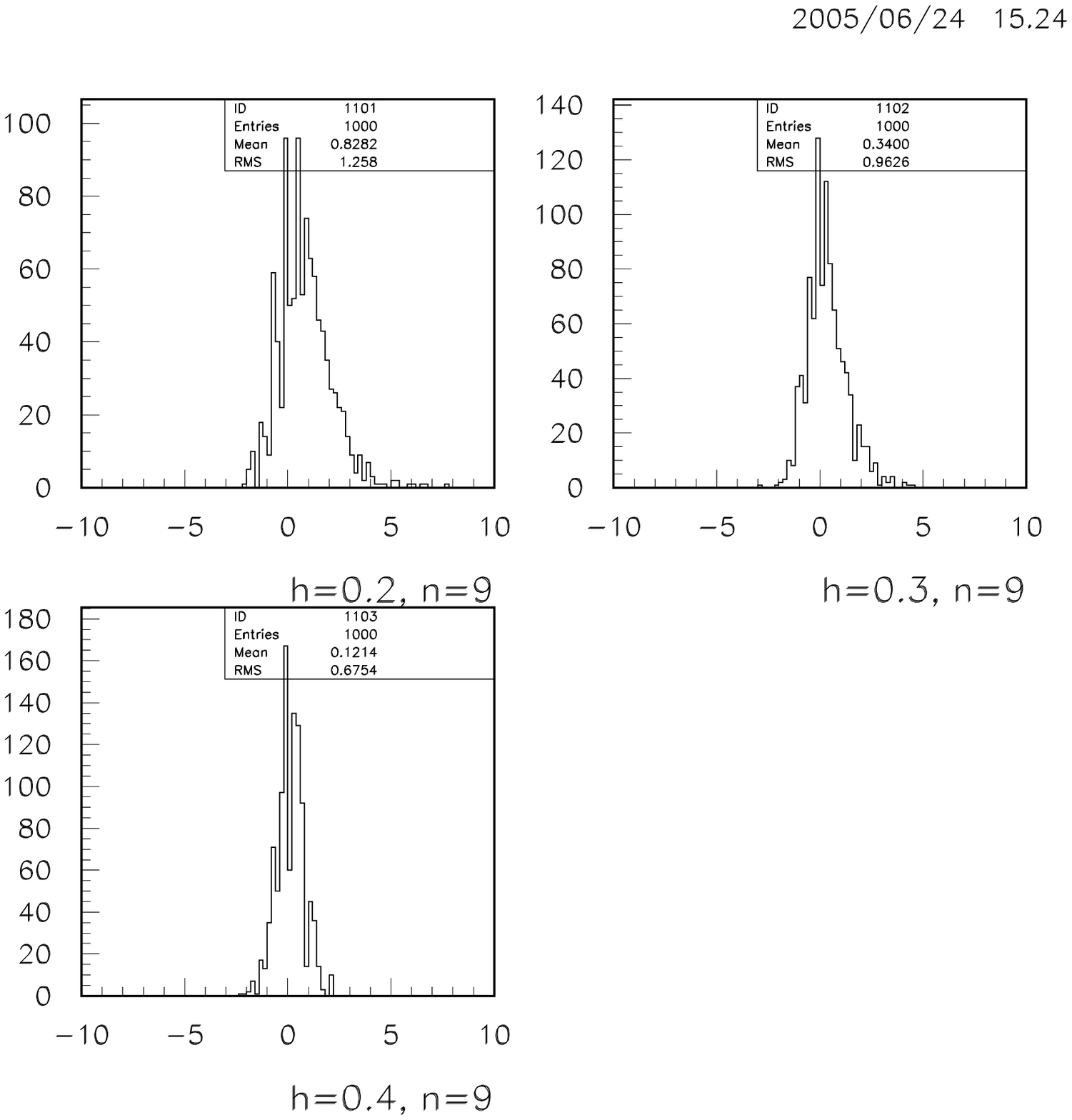}
\caption
{The distribution of ${\cal NLLR}$ as a function of the smoothing 
parameter $h=0.2,0.3,0.4$ for a dataset $n=9$ generated to be uniform in the hypercube.
\label{null9}}
\end{figure}

\begin{figure}[tbh!]
\includegraphics[width=\textwidth]{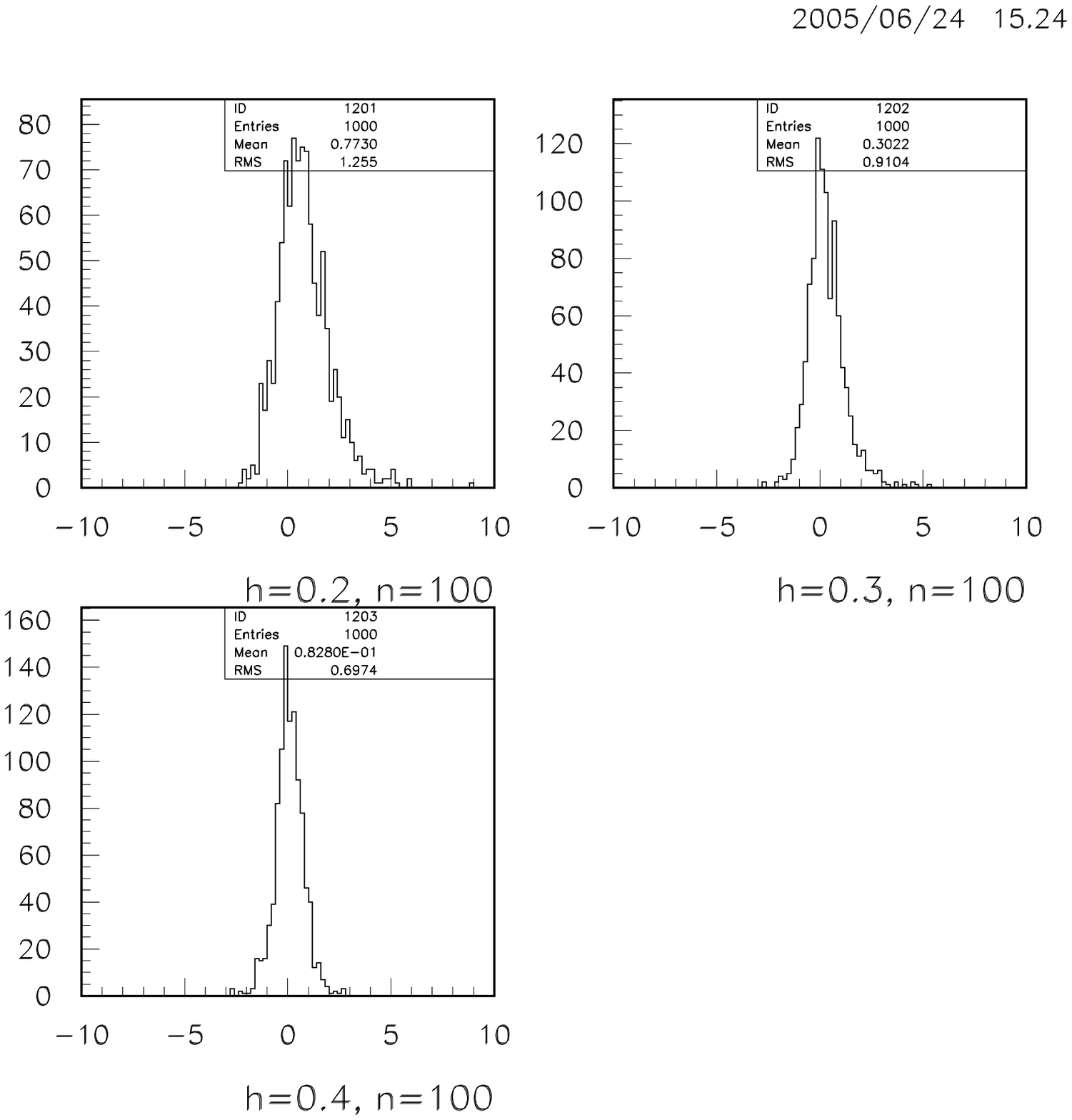}
\caption
{The distribution of ${\cal NLLR}$ as a function of the smoothing 
parameter $h=0.2,0.3,0.4$ for a dataset $n=100$ generated to be uniform in the hypercube.
\label{null100}}
\end{figure}

Figure~\ref{hyper} shows the transformed co-ordinates in hypercube space.
We then proceed to work out the negative log-likelihood ratio ${\cal NLLR}$
for this
configuration with the ``smoothing parameter $h$'' set to three
different values $h=0.2,0.3$ and $0.4$. We study the behavior of the
${\cal NLLR}$ for the null hypothesis (i.e. n=9 events distributed
uniformly in hypercube space) for a 1000 such experiments. We repeat
this for a dataset of $n=100$ as well to study the effect of the small
data sample on our goodness of fit measure.
Figure~\ref{null9} shows the distribution of the ${\cal NLLR}$ for the 
three different values of $h$ for a data set size $n=9$.
Figure~\ref{null100} shows the distribution of the ${\cal NLLR}$ for the 
three different values of $h$ for a data set size $n=100$.
Table~\ref{summary} summarizes the observed ${\cal NLLR}$ for our
dataset as a function of $h$. The mean and sigma of the null
hypothesis histograms are also shown as well as the probability that
the observed ${\cal NLLR}$ is exceeded for both the $n=9$ null
hypothesis and an $n=100$ null hypothesis. The latter is run to test
the sensitivity of the results to the small data sample.
\begin{table}
\caption[]{Summary of results\label{summary}}
\begin{center}
\begin{tabular}{|c|c|c|c|c|c|c|c|}
\hline
Smoothing  & Dataset & n=9 & n=9 & n=9 Prob.& n=100 & n=100 & n=100 Prob. \\
  parameter $h$   & ${\cal NLLR}$ & $\mu$ & $\sigma$ & to exceed 
                                & $\mu$ & $\sigma$ & to exceed  \\
\hline
0.2 & 5.36 & 0.82 & 1.26 & 0.5\%& 0.77 & 1.255 & 0.3\%\\
\hline
0.3 & 5.84 & 0.34 & 0.96 & $<0.1\%$ & 0.30 & 0.91 & $<0.1\%$ \\
\hline
0.4 & 1.77 & 0.12 & 0.67 & 1.0\% & 0.083 & 0.697 & 1.1\%\\
\hline
\hline
\end{tabular}
\end{center}
\end{table}
\subsection{Comments}
The observed data is a bad fit to the model.
We have managed not only obtain a goodness of fit for the 
problem (made extreme by the sparsity of data), but also to show that
the method gives reliable results for a variety of smoothing
parameters. The method is also robust with respect to the data size
$n$. We see that as we increase the smoothing parameter to 0.4, we
begin to increase the chance of fitting. When $h=1.0$, everything
will fit. A smoothing parameter of $h=0.2$ or $0.3$ gives reliable
results.  The probability to exceed the observed ${\cal NLLR}$ is
estimated from the histograms with 1000 experiments. We can improve
the accuracy of this by running more Monte Carlo statistics.


\begin{thebibliography}{9}
\bibitem{power} Empirically we found that a power of the order of -0.6 is
needed to provide sufficiently large smoothing factors for large values of time.
We can in principle optimize this smoothing function further, 
but have not done so.
\end{thebibliography}

\end{document}